\documentclass[10pt,twocolumn]{article}

\usepackage[a4paper,margin=1.8cm]{geometry}
\usepackage{graphicx}
\usepackage{amsmath}
\usepackage{amssymb}
\usepackage{physics}
\usepackage{mathtools}

\usepackage{tikz}
\usetikzlibrary{arrows.meta,decorations.pathreplacing}

\usepackage[dvipsnames]{xcolor}
\usepackage{booktabs}
\usepackage{placeins}

\usepackage{caption}
\usepackage{subcaption}
\usepackage[super,sort&compress]{natbib}

\usepackage{hyperref}
\usepackage{xr-hyper}
\hypersetup{hidelinks}

\usepackage{comment}

\definecolor{figureblue}{RGB}{80,100,170}
\definecolor{figurered}{RGB}{170,85,105}
\definecolor{figureorange}{RGB}{200,130,50}

\usepackage{ragged2e}

\usepackage{authblk}

\title{Dynamical regimes of QAOA gradient response}

\author[1]{Zarin Shakibaei}
\author[2]{Alexander Schnell}

\affil[1]{Technische Universit\"at Berlin, Institute of Mathematics, 10623 Berlin, Germany}
\affil[2]{Technische Universit\"at Berlin, Institute for Physics and Astronomy, 10623 Berlin, Germany}

\date{\today}

\begin{document}

\maketitle

\begin{abstract}

Characterizing the trainability of the Quantum Approximate Optimization Algorithm (QAOA) requires understanding how its gradient landscape changes across circuit parameters and problem size. Here we introduce a dynamical representation of QAOA parameter space based on a norm-weighted layer strength and a cost–mixer ratio, separating the overall scale of the evolution from the relative contribution of the two generators. Using exact-state simulations of MaxCut, we find that the gradient landscape exhibits a coarse organization in these dynamical variables that persists across circuit depth and schedule structure, while finer interference patterns remain schedule dependent. Near-optimal solutions occupy an extended intermediate, balanced-to-cost-biased dynamical regime rather than regions of largest local gradients. As system size increases, these favorable regions remain extended in dynamical coordinates while their preimages in the native QAOA angles become increasingly compressed. In the weak-drive regime, we derive an analytical prediction for the onset of  near-optimal sampling, explicitly separating the contributions of cost–mixer ratio, circuit depth, and problem-dependent spectral properties. These results distinguish the persistence of useful QAOA dynamics from their accessibility in the native parameterization and provide a dynamical framework for interpreting gradient response and trainability in structured QAOA circuits. 
\end{abstract}

\section{Introduction}
\label{sec:introduction}
Variational quantum algorithms optimize parametrized quantum circuits by using classical feedback from measured observables. Their practical performance therefore depends on whether parameter variations produce resolvable changes in the objective. This question has motivated extensive work on variational trainability, including barren plateaus~\cite{McClean2018,Cerezo2021Review}, expressibility~\cite{Sim2019,Holmes2022}, cost-function structure~\cite{Cerezo2021CostDependent}, and the geometry of optimization landscapes. Yet for structured algorithms such as the Quantum Approximate Optimization Algorithm (QAOA)~\cite{Farhi2014}, the connection
between variational parameters and the underlying circuit dynamics remains only partly understood.

The native parameters of QAOA provide the standard control coordinates, but they do not separately describe the overall scale of the applied dynamics and the relative contribution of the two generators. This distinction becomes particularly important when comparing different instances or
system sizes, for which the characteristic scales of the generators can change. Previous studies have shown that optimized QAOA parameters can exhibit regularity, concentration, and transferability across related instances and circuit depths~\cite{Zhou2020QAOAPerformance,Akshay2021}, and have connected QAOA performance to schedule and control  structure and controllability~\cite{CamposVenuti2021OptimalControl,Brady2021OptimalProtocols, Sack2021QuantumAnnealing,Arezzo2026DigitalControllability}. These observations motivate asking whether a simpler dynamical organization underlies the native parameter landscape.

Here, we introduce a strength--imbalance representation of QAOA parameter space that separates the total norm-weighted action of a layer from the relative contribution of the cost and mixer generators. This representation provides a common dynamical language for comparing parameter landscapes across circuit depths, schedules, instances, and system sizes, while
distinguishing changes in the underlying dynamics from those associated with the native parameterization.

Using exact-state simulations of MaxCut, we find that broad responsive and suppressed regions remain recognizable as circuit depth and layerwise schedule structure are varied, indicating that the coarse landscape is governed primarily by strength and cost--mixer imbalance, while depth and schedule details mainly reshape the finer interference pattern.

This distinction also proves relevant to optimization performance. The probability of sampling near-optimal solutions is organized in the strength–imbalance plane, with favorable parameter regions concentrated at intermediate strength and balanced-to-cost-biased imbalance. Across the uniform and nonuniform schedules and depths considered, this region remain localized in broadly similar portions of the dynamical plane, although their detailed structure and preferred location remain schedule dependent.

Beyond identifying this favorable regime, we show that its weak-drive onset can be predicted analytically. The resulting boundary separates a characteristic dependence on cost--mixer imbalance and circuit depth from an instance--dependent contribution determined by the cost spectrum. 

The dynamical representation further changes the interpretation of system-size scaling. We find that regions supporting responsive gradients and near-optimal solutions can remain extended in dynamical coordinates while their representations in the native QAOA parameters become substantially compressed at larger system sizes. This separates the persistence of useful QAOA dynamics from their accessibility.

Finally, we connect this regime picture to the underlying state dynamics. At strong drive, unweighted and real-weighted cost Hamiltonians exhibit distinct recurrence and state-spreading behavior, linking the landscape structure to phenomena familiar from periodically driven quantum systems
~\cite{DAlessio2014,Bukov2015,Kuwahara2016}. 
Entanglement and participation diagnostics further reveal a strength--imbalance-dependent crossover in state spreading, complementing earlier links between entanglement and variational trainability~\cite{Marrero2021Entanglment}. A state-dependent trace-speed bound provides a complementary geometric perspective by separating dynamically allowed state motion from the gradient response actually realized. Together, these results support a unified picture in which QAOA trainability is organized by the dynamical regime occupied by the circuit, rather than by native parameter magnitude or gradient size in isolation.


\section{Results}
\label{sec:results}

\subsection{Strength–imbalance framework for QAOA dynamics}
\label{subsec:drive-imbalance-framework}

At depth $p$, QAOA alternates evolution under a cost Hamiltonian $H_c$ and a mixer Hamiltonian $H_m$, with native parameters $\gamma_\ell,\beta_\ell\geq0$,
\begin{equation}
U_p
=
\prod_{\ell=1}^{p}
e^{-i \beta_\ell H_m}
e^{-i \gamma_\ell H_c}.
\label{eq:propagator}
\end{equation}
We interpret $\gamma_\ell$ and $\beta_\ell$ as the corresponding effective evolution times; the controlled-dynamics formulation is given in Methods, Sec.~\ref{sec:formulation}.

For general time-dependent quantum systems with Hamiltonian $H(t)$, a natural measure of the total dynamical action is
\begin{equation}
S^\mathrm{tot}(T) := \int_0^T \|H(t)\|\, \mathrm{d}t,
\label{eq:cumulative-strength}
\end{equation}
where $\|\cdot\|$ denotes the spectral norm. This quantity plays a central role in the convergence of Magnus expansions~\cite{Magnus1954,Blanes2009}, the accuracy of product-formula approximations~\cite{Suzuki1992,Berry2007,Childs2021}, and finite-time controllability~\cite{Jurdjevic1997}.
Motivated by this viewpoint, we describe QAOA dynamics using two quantities: the total dynamical action applied within each layer and the relative contribution of the cost and mixer generators. Together, these quantities determine the scale and structure of the evolution.

At the layer level, the same norm-weighted combination also bounds the trace-distance displacement of the state,
\[
D(\rho_{\ell-1},\rho_\ell)
\le
\gamma_\ell\|H_c\|+\beta_\ell\|H_m\|,
\]
as shown in Methods, Sec.~\ref{sec:structure}. This identifies the norm-weighted cost and mixer actions as natural dynamical scales for a QAOA layer.

For a single layer, we define
\begin{equation}
S_\ell := \gamma_\ell \|H_c\| + \beta_\ell \|H_m\|,
\label{eq:layer-strength}
\end{equation}
which serves as a convenient measure of the applied dynamical drive. While this determines the magnitude of the evolution it does not capture the structure. Therefore we define the cost-mixer ratio
\begin{equation}
\alpha_\ell :=
\frac{\gamma_\ell\|H_c\|}
{\beta_\ell\|H_m\|}.
\label{eq:imbalance}
\end{equation}
This dimensionless quantity measures the relative weight of the generators. When $\alpha_\ell \ll 1$ or $\alpha_\ell \gg 1$, the dynamics becomes dominated by a single generator. When $\alpha_\ell \approx 1$, both generators act on comparable scales, enabling noncommutative dynamics. Thus, $\alpha_\ell$ quantifies generator competition.

\subsubsection{Structure of effective dynamics}
\label{subsec:bch-depth}

Building on the dynamical formulation above, we analyze how the layered structure of QAOA determines the effective generator and how the full parameter schedule shapes the resulting dynamics. This distinguishes algebraic expressivity, set by the generators, from dynamical realization, which depends on the total evolution time and how it is distributed across layers.

We begin at the level of a single layer. By writing
\begin{equation}
U_\ell = e^{-i\beta_\ell H_m} e^{-i\gamma_\ell H_c},
\label{eq:layer-unitary}
\end{equation}
the Baker--Campbell--Hausdorff (BCH) formula yields
\begin{equation}
\log U_\ell
=
-i(\gamma_\ell H_c + \beta_\ell H_m)
- \frac{1}{2}\gamma_\ell \beta_\ell [H_m, H_c]
+ O(\tau_\ell^3).
\end{equation}
The leading term generates evolution under a weighted sum of $H_c$ and $H_m$, which corresponds to the commutative approximation. The next term captures the effect of their noncommutativity, while higher-order contributions involve nested commutators that are suppressed for small layer durations $\tau_\ell$, defined in Eq. \eqref{eq:time_partition}. All terms remain within the dynamical Lie algebra $\mathfrak{g}$.

For the full circuit, we have
\begin{equation}
U_p = \prod_{\ell=1}^p U_\ell = e^{\Omega_p},
\end{equation}
where $\Omega_p$ depends on the full parameter schedule and the ordering of layers. This corresponds to an evolution under a piecewise constant Hamiltonian alternating between $H_c$ and $H_m$. Although \(\Omega_p\) can in principle be obtained by repeated applications of the Baker--Campbell--Hausdorff (BCH) formula, the layered QAOA evolution is more naturally described by the Magnus expansion~\cite{Magnus1954,Blanes2009}. More generally, the Magnus expansion expresses the solution of $\dot U(t)=A(t)U(t)$ as a single exponential $U(t)=e^{\Omega(t)}$, where $\Omega(t)$ is given by a series of time-ordered integrals involving $A(t)$ and its nested commutators. In the present case, $A(t)=-iH(t)$, with $H(t)$ piecewise constant and alternating between $H_c$ and $H_m$.

For a depth-\(p\) QAOA circuit, the expansion takes the form
\begin{equation}
\Omega_p = \Omega_p^{(1)} + \Omega_p^{(2)} + \Omega_p^{(3)} + \cdots,
\end{equation}
where the terms are organized by order in the layer durations. The first-order term collects all layer contributions,
\begin{equation}
\Omega_p^{(1)} = -i \sum_{\ell=1}^p (\gamma_\ell H_c + \beta_\ell H_m),
\end{equation}
and does not depend on the ordering of layers.

Noncommutative effects first appear at second order, which contains both intra-layer and inter-layer contributions,
\begin{align}
\Omega_p^{(2)}
&=
-\frac{1}{2} \sum_{\ell=1}^p \gamma_\ell \beta_\ell [H_m, H_c] 
+
\frac{1}{2} \sum_{\ell < k} \Big( \beta_\ell \gamma_k - \gamma_\ell \beta_k \Big)[H_m, H_c].
\label{eq:Magnus-2nd}
\end{align}
The first term depends on parameters within a single layer, while the second captures how different layers combine and therefore reflects the ordering of the circuit. Intra-layer terms scale as $O(\gamma_\ell \beta_\ell)$, which is bounded by $O(\tau_\ell^2)$, while inter-layer terms scale as $O(\tau_\ell \tau_k)$. In both cases, the magnitude of these contributions is set by products of layer durations.

At fixed Magnus order $r$, contributions involve products of $r$-layer-dependent parameters and scale as $O(\tau_{\ell_1}\cdots\tau_{\ell_r})$. To characterize the scaling of higher-order contributions, we fix the total evolution time \(T\) defined in Eq.~\eqref{eq:total_time} and consider a representative regime in which layer durations are comparable, \(\tau_\ell \sim T/p\). Under this scaling, each \(r\)th-order term scales as \((T/p)^r\).

However, the number of such contributions depends on how many distinct layers are involved. Terms involving a single layer appear in $O(p)$ combinations and therefore scale as $O(T^r/p^{r-1})$, becoming small for large $p$. In contrast, terms involving $r$ distinct layers arise from $\binom{p}{r}$ combinations. For fixed $r$ and large $p$, $\binom{p}{r}\sim p^r/r!$, so these contributions collectively remain generically of order $O(T^r)$. Thus, contributions localized within individual layers are suppressed under this scaling, while contributions involving $r$ distinct layers can remain of order $O(T^r)$. The overall scale of the expansion remains controlled by the total evolution time $T$, but its structure changes.

As the number of layers $p$ increases while the total evolution time remains fixed, more noncommutative contributions appear, but their individual magnitudes decrease as the time is distributed across layers. The relative structure of the effective generator therefore shifts toward contributions involving multiple layers. This structural behavior will be made more precise in Sec.~\ref{subsec:imbalance-bch-structure}, where we provide a quantitative description of how these contributions are weighted.

We now express this noncommutative structure directly in the strength--imbalance coordinates.

\subsubsection{\texorpdfstring{$(S_\ell,\alpha_\ell)$ and noncommutative structure}{Strength, imbalance, and noncommutative structure}}
\label{subsec:imbalance-bch-structure}

To make the role of the layer strength and cost--mixer ratio clear, we express the Magnus expansion of the QAOA circuit in terms of \(S_\ell\) and \(\alpha_\ell\)~\cite{Magnus1954,Blanes2009}. At second order, where noncommutative effects first appear, the dependence on strength and cost--mixer ratio can be made explicit. Using that
\begin{equation}
\gamma_\ell=
\frac{\alpha_\ell}{1+\alpha_\ell}\frac{S_\ell}{\|H_c\|},
\qquad
\beta_\ell=
\frac{1}{1+\alpha_\ell}\frac{S_\ell}{\|H_m\|},
\label{eq:strength-imbalance-map}
\end{equation}
substituting these expressions into the second-order term of Eq.~\eqref{eq:Magnus-2nd} gives
\begin{align}
\Omega_p^{(2)}
&=
-\frac{1}{2}
\sum_{\ell=1}^p
\frac{S_\ell^2}{\|H_c\|\,\|H_m\|}
\frac{\alpha_\ell}{(1+\alpha_\ell)^2}
[H_m,H_c]
\nonumber\\
&\quad+
\frac{1}{2}
\sum_{\ell<k}
\frac{S_\ell S_k}{\|H_c\|\,\|H_m\|}
\frac{\alpha_k-\alpha_\ell}
{(1+\alpha_\ell)(1+\alpha_k)}
[H_m,H_c].
\label{eq:omega2-strength-imbalance}
\end{align}

The first line of Eq.~\eqref{eq:omega2-strength-imbalance} is the layer-local contribution. Its factor, \(\alpha_\ell/(1+\alpha_\ell)^2\), is maximized at \(\alpha_\ell=1\) and vanishes in the strongly imbalanced limits, as \(\alpha_\ell\to0\) or
\(\alpha_\ell\to\infty\). Thus, extreme cost--mixer imbalances suppress the local commutator within a layer, while balanced driving enhances it. The second line of Eq.~\eqref{eq:omega2-strength-imbalance} represents the inter-layer contribution. Its ratio dependence is controlled by differences between layer ratios, \(\alpha_k-\alpha_\ell\), and therefore vanishes at second order for schedules with an equally constant cost--mixer ratio. For nonuniform schedules, these inter-layer terms are nonzero and can accumulate over many layer pairs.

At higher orders, the same reparametrization produces prefactors combining powers of the relevant layer strengths with functions of the corresponding cost--mixer ratios. Layer-local mixed terms are suppressed in strongly
imbalanced limits, while inter-layer terms depend on how the ratios vary across the schedule. Thus, \(S_\ell\) and \(\alpha_\ell\) provide a compact way to organize both local cost--mixer structure and schedule-dependent inter-layer
structure.

As we will see later in Sec.~\ref{subsec:numerics-solution-quality}, the same strength--imbalance factor appears in the perturbative onset of the near-optimal sampling probability.

This commutator structure connects to the gradient response through the parameter generators derived in Sec.~\ref{sec:structure}, for which
\begin{equation}
\partial_\theta C
=
i\,\mathrm{Tr}\!\left(\rho_\theta [G_\theta,H_c]\right).
\end{equation}
For \(\theta=\gamma_\ell\),
\begin{equation}
\begin{aligned}
G_{\gamma_\ell}
&= U_{>\gamma_\ell}H_cU_{>\gamma_\ell}^\dagger =
e^{\Omega_{>\gamma_\ell}}
H_c
e^{-\Omega_{>\gamma_\ell}}
\\
&=
H_c
+
[\Omega_{>\gamma_\ell},H_c] +
\\
&\quad
\frac{1}{2}
[\Omega_{>\gamma_\ell},
 [\Omega_{>\gamma_\ell},H_c]]
+\cdots ,
\end{aligned}
\end{equation}
where \(U_{>\gamma_\ell}\) denotes the circuit segment following the parameter \(\gamma_\ell\), and \(\Omega_{>\gamma_\ell}\) is the corresponding Magnus generator. Hence, the \(S_\ell\)- and \(\alpha_\ell\)-dependent coefficients of the Magnus generator directly weight the commutator corrections entering the gradient response.


\subsection{Dynamical Regimes of Gradient Response}
\label{sec:regimes}

The strength--imbalance framework introduced above provides a dynamical classification of the QAOA evolution in terms of the layer strength $S_\ell$
and the cost--mixer ratio $\alpha_\ell$. These variables identify qualitatively distinct dynamical regimes in which different contributions to the evolution and gradient response become relevant. Although the classification is defined at the level of individual layers, its effects accumulate across the circuit through the time-ordered evolution.

\subsubsection{Definition of dynamical regimes}
We distinguish four qualitative regimes in the $(S_\ell,\alpha_\ell)$ plane, summarized schematically in Fig.~\ref{fig:regimes}.

\paragraph{Weak-drive regime.}
For
\[
S_\ell \ll 1,
\]
the layer unitary remains close to the identity and higher-order terms in the Magnus expansion are perturbatively suppressed. The state therefore undergoes only limited displacement during the layer, restricting the variation of the objective and leading to small gradients. This is the geometric suppression mechanism identified in Sec.~\ref{sec:structure}: insufficient applied evolution produces weak parameter sensitivity.

\begin{figure}[t]
\centering
\begin{tikzpicture}[scale=1, every node/.style={font=\footnotesize}]


\fill[gray!10] (0,0) rectangle (1.7,5);          
\fill[gray!6]  (1.7,3.4) rectangle (4.2,5);     
\fill[gray!10] (1.7,1.6) rectangle (4.2,3.4);   
\fill[gray!6]  (1.7,0) rectangle (4.2,1.6);     
\fill[gray!18] (4.2,0) rectangle (7.1,5);       


\begin{scope}
  \clip (0,3.4) rectangle (7.1,5);
  \foreach \x in {-4,-3.7,...,8} {
    \draw[gray!55, line width=0.2pt]
      (\x,3.35) -- ++(2.2,2.2);
  }
\end{scope}

\begin{scope}
  \clip (0,0) rectangle (7.1,1.6);
  \foreach \x in {-1,-0.7,...,10} {
    \draw[gray!55, line width=0.2pt]
      (\x,1.65) -- ++(2.2,-2.2);
  }
\end{scope}

\draw[dashed, gray!55] (1.7,0) -- (1.7,5);
\draw[dashed, gray!55] (4.2,0) -- (4.2,5);

\draw[dashed, gray!55] (0,1.6) -- (4.2,1.6);
\draw[dashed, gray!55] (0,3.4) -- (4.2,3.4);

\draw[-{Latex[length=2mm,width=1.5mm]}, semithick]
  (0,0) -- (7.45,0);
\draw[-{Latex[length=2mm,width=1.5mm]}, semithick]
  (0,0) -- (0,5.35);

\node[below] at (7.45,0) {$S_\ell$};
\node[left]  at (0,5.35) {$\alpha_\ell$};

\node[below] at (0.85,0) {$S_\ell \ll 1$};
\node[below] at (2.95,0) {intermediate $S_\ell$};
\node[below] at (5.65,0) {large $S_\ell$};

\node[left] at (0,4.2) {$\alpha_\ell \gg 1$};
\node[left] at (0,2.5) {$\alpha_\ell \simeq 1$};
\node[left] at (0,0.8) {$\alpha_\ell \ll 1$};

\node[align=center] at (0.85,2.5)
  {Weak drive};

\node[align=center] at (2.95,4.2)
  {Cost dominated};

\node[align=center] at (2.95,2.5)
  {Balanced\\cost--mixer drive};

\node[align=center] at (2.95,0.8)
  {Mixer dominated};

\node[align=center] at (5.65,2.5)
  {Strong drive\\higher-order dynamics\\
   {\scriptsize potential mixing}};

\end{tikzpicture}

\caption{\justifying
    \textbf{Schematic dynamical regimes in the $(S_\ell,\alpha_\ell)$ plane.} Weak drive, balanced and strongly imbalanced intermediate-drive regimes, and strong drive are indicated schematically. Hatched regions denote single-generator-dominated limits. Dashed lines indicate schematic, problem-dependent crossovers rather than sharp boundaries.
}
\label{fig:regimes}
\end{figure}


\paragraph{Balanced cost--mixer regime.}
At intermediate strength and
\[
\alpha_\ell \simeq 1,
\]
the cost and mixer generators contribute on comparable norm-weighted scales. As shown in
Eq.~\eqref{eq:omega2-strength-imbalance}, the intra-layer second-order commutator contribution is maximal for fixed \(S_\ell\). The detailed gradient response nevertheless remains schedule dependent through inter-layer and higher-order contributions.

\paragraph{Strongly imbalanced regime.}
For
\[
\alpha_\ell \ll 1
\qquad \text{or} \qquad
\alpha_\ell \gg 1,
\]
one generator dominates the evolution. The same mixed-commutator factor satisfies
\[
\frac{\alpha_\ell}{(1+\alpha_\ell)^2}\rightarrow 0
\]
in either limit, suppressing local competition between $H_c$ and $H_m$. The resulting sensitivity can therefore become strongly anisotropic, with parameter variations associated with the subdominant generator producing a comparatively weak response. This imbalance-induced suppression is distinct from the weak-drive mechanism because it can occur even when the overall applied strength is appreciable.

\paragraph{Strong-drive regime.}
At large $S_\ell$, higher-order terms in the Magnus expansion become significant. 
In driven many-body systems, such nonperturbative dynamics can lead to strong state mixing and thermalization~\cite{DAlessio2014,Bukov2015}. If comparable mixing develops in QAOA, the local sensitivity of the cost expectation to parameter variations may be suppressed. However, strong drive alone does not imply mixing, and recurrence, spectral structure, and circuit details can preserve substantial gradient response.

\medskip

\noindent

This regime picture has a natural connection to periodically driven quantum systems. QAOA alternates evolution under two noncommuting generators and, for uniform schedules, repeats the same elementary layer across the circuit, giving the dynamics a stroboscopic structure closely related to Floquet evolution while remaining a finite-depth control protocol rather than a long-time evolution. Related regime structures arise in driven quantum systems and quantum chaos~\cite{Haake2010}. A prominent example is the quantum kicked rotor~\cite{Chirikov1979,Casati1979,Santhanam2022}, whose Floquet map alternates between free rotation and impulsive driving and exhibits qualitatively different behavior as the kicking strength is varied. More generally, weak or high-frequency driving can often be treated perturbatively through a Floquet--Magnus expansion~\cite{Bukov2015,Eckardt2015,Kuwahara2016}, whereas at stronger drive this description can break down, allowing heating~\cite{Eckardt2017}, mixing~\cite{DAlessio2014,Bukov2015}, and more complex dynamical behavior to emerge.

\subsubsection{System-size scaling of native accessibility}
\label{subsec:size-scaling}

We next consider how the strength--imbalance coordinates and their mapping to the native parameters change with system size $n$. The dependence on $n$ enters through the generator norms. For the Hamiltonians considered here, $\|H_m\|=n$. For unweighted MaxCut, $\|H_c\|=C_{\max}$, and the maximum cut satisfies $|E|/2\leq C_{\max}\leq |E|$. The cost-Hamiltonian norm therefore grows proportionally to the number of edges: linearly with $n$ for sparse graph families with $|E|\propto n$, and quadratically with $n$ for dense
families with $|E|\propto n^2$.

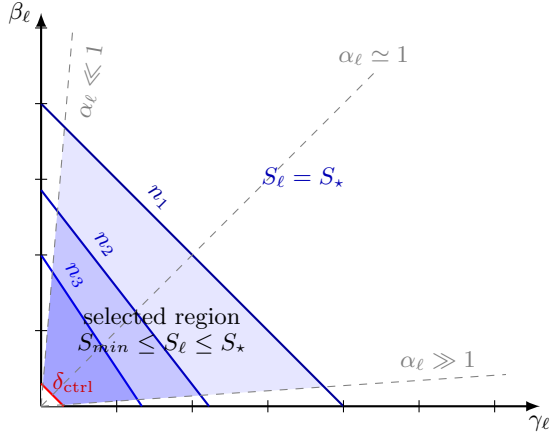
\begin{figure}[!t]
\centering
\begin{tikzpicture}[scale=1.,>=latex]

  \draw[->, thick] (0,0) -- (6.6,0) node[below] {$\gamma_\ell$};
  \draw[->, thick] (0,0) -- (0,5.2) node[left] {$\beta_\ell$};

  \begin{scope}
    \clip (0,0) -- (6.15,0.42) -- (0.42,5.0) -- cycle;
    \fill[blue!10]
      (0,0) -- (4.0,0) -- (0,4.0) -- cycle;
  \end{scope}

  \begin{scope}
    \clip (0,0) -- (6.15,0.42) -- (0.42,5.0) -- cycle;
    \fill[blue!22]
      (0,0) -- ({4.0/1.8},0) -- (0,{4.0/1.4}) -- cycle;
  \end{scope}

  \begin{scope}
    \clip (0,0) -- (6.15,0.42) -- (0.42,5.0) -- cycle;
    \fill[blue!36]
      (0,0) -- ({4.0/3.0},0) -- (0,{4.0/2.0}) -- cycle;
  \end{scope}

    \fill[white]
      (0,0) -- (0.3,0) -- (0,0.3) -- cycle;
    
    \draw[thick, red]
      (0.3,0) -- (0,0.3);

  \draw[thick, blue!60!black]
    (4.0,0) -- (0,4.0)
    node[pos=0.65, sloped, above=0.25pt] {$n_1$};

  \draw[thick, blue!75!black]
    ({4.0/1.8},0) -- (0,{4.0/1.4})
    node[pos=0.7, sloped, above=0.25pt] {$n_2$};

  \draw[thick, blue!90!black]
    ({4.0/3.0},0) -- (0,{4.0/2.0})
    node[pos=0.8, sloped, above=0.25pt] {$n_3$};

    \node[red!80!black]
      at (0.43,0.32)
      {$\delta_{\rm ctrl}$};

  \draw[dashed, gray] (0,0) -- (4.4,4.4)
    node[above] {$\alpha_\ell \simeq1$};

  \draw[dashed, gray] (0,0) -- (6.15,0.42);
  \node[gray] at (5.25,0.6) {$\alpha_\ell \gg 1$};

  \draw[dashed, gray] (0,0) -- (0.42,5.0);
  \node[gray, rotate=80] at (0.65,4.3) {$\alpha_\ell \ll 1$};

  \node[align=center] at (1.6,1)
    {\small selected region\\[-1pt]\small $ S_{min} \le S_\ell \le S_\star$};

  \node[align=center, blue!70!black] at (3.47,3.08) 
    {\small \\[-1pt]\small $S_\ell = S_\star$};

  \foreach \x in {1,2,3,4,5,6}
    \draw (\x,0.08) -- (\x,-0.08);
  \foreach \y in {1,2,3,4,5}
    \draw (0.08,\y) -- (-0.08,\y);

\end{tikzpicture}

\caption{\justifying
    \textbf{Native-coordinate geometry and contraction with system size.}
    A selected region of the strength--imbalance plane, represented schematically by a finite strength interval
    $S_{\min}\leq S_\ell\leq S_\star$ and restricted away from strongly imbalanced limits, maps to a finite region of the native $(\gamma_\ell,\beta_\ell)$ plane. Constant strength defines straight boundaries, whereas constant imbalance defines rays from the origin. As the generator norms grow with system size ($n_1<n_2<n_3$), the native preimage of the same dynamical region contracts with characteristic scales set by $\|H_C\|^{-1}$ and $\|H_M\|^{-1}$. The imbalance rays are schematic; their slopes depend on $\|H_C\|/\|H_M\|$ and therefore generally vary with
    system size. The red dashed line denotes an illustrative finite control scale $\delta_{\rm ctrl}$; no specific hardware resolution is assumed.
    }
\label{fig:size-dependent-window}
\end{figure}

Maintaining the same region of the strength--imbalance plane as the system grows requires the native parameters to decrease with the corresponding generator norms. From the inverse map in Eq.~\eqref{eq:strength-imbalance-map}, a bounded region of the $(S_\ell,\alpha_\ell)$ plane therefore maps to a progressively smaller region of the native $(\gamma_\ell,\beta_\ell)$ plane as $n$ increases. This geometry is illustrated schematically in Fig.~\ref{fig:size-dependent-window}.

The contraction can be quantified directly from the Jacobian of the coordinate transformation,
\begin{equation}
\left|
\frac{\partial(\gamma_\ell,\beta_\ell)}
     {\partial(S_\ell,\alpha_\ell)}
\right|
=
\frac{S_\ell}
{\|H_c\|\,\|H_m\|\,(1+\alpha_\ell)^2}.
\label{eq:size-jacobian}
\end{equation}
Thus, for any fixed bounded region in the strength--imbalance plane, the corresponding native area scales as
\[
\operatorname{Area}_{\gamma,\beta}
=
O\!\left(
\frac{1}{\|H_c\|\,\|H_m\|}
\right).
\]
For sparse MaxCut families this gives
$\Delta\gamma=O(n^{-1})$ and $\Delta\beta=O(n^{-1})$, whereas for dense
families $\Delta\gamma=O(n^{-2})$ and $\Delta\beta=O(n^{-1})$. Thus, for dense graphs, the native preimage contracts more rapidly along the $\gamma_\ell$ axis than along the $\beta_\ell$ axis as $\|H_c\|$ grows faster than $\|H_m\|$.

This contraction is purely geometric and places no constraint on the magnitude of the gradients within that region. This contraction is geometric: it does not, by itself, imply that the gradients within that region become smaller. The same reasoning applies when targeting specific regions of interest, such as near-optimal solution region or gradient-responsive regions; their corresponding native preimages are examined in Supplementary Sec.~I\,H. 

\begin{figure*}[t]
    \centering
    \includegraphics[width=0.85\textwidth]
    {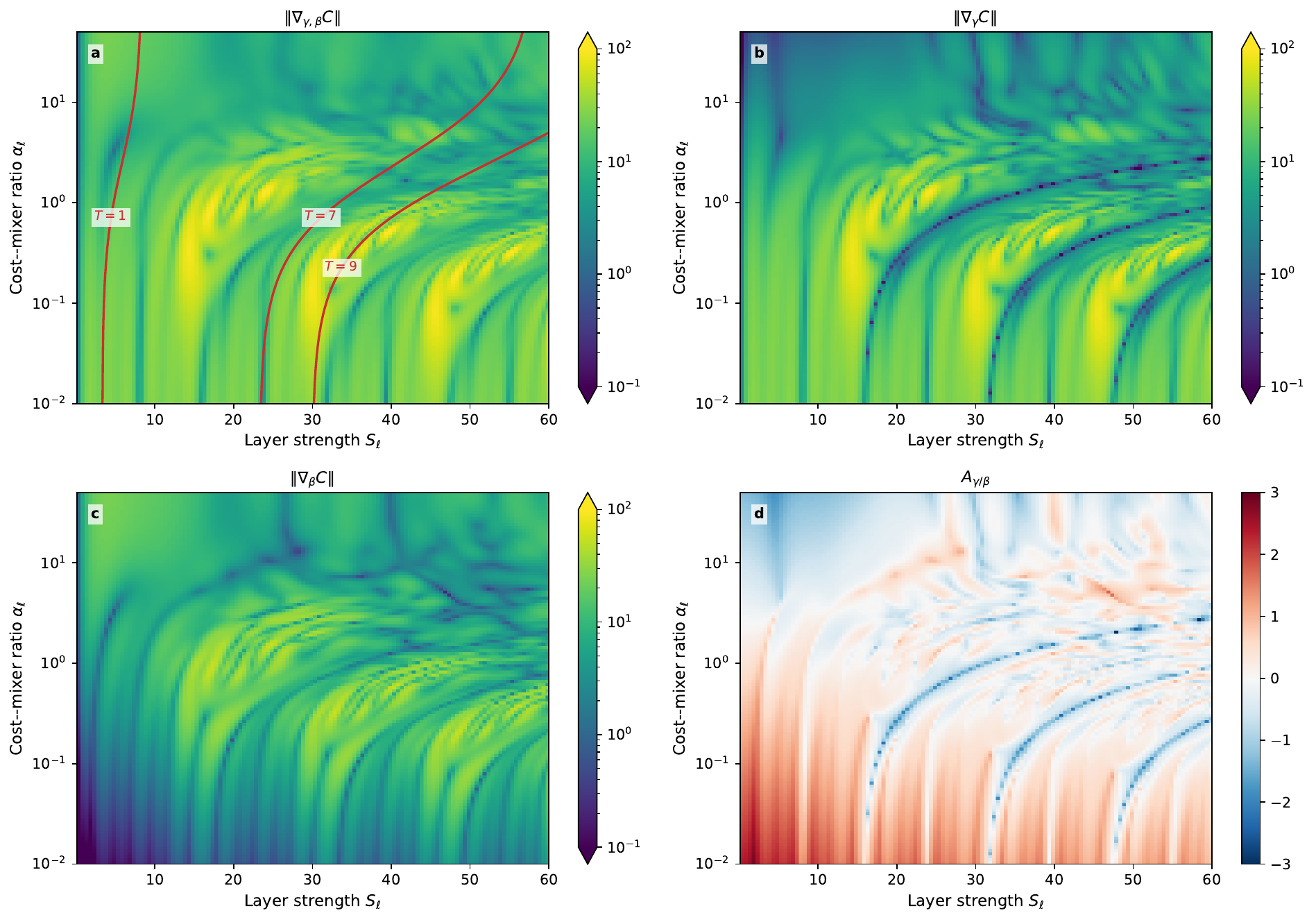}
    \caption{\justifying \textbf{Dynamical organization of the native-gradient response.}
    Heatmaps over the $(S_\ell,\alpha_\ell)$ plane for a representative dense, unweighted MaxCut instance with $n=10$, depth $p=3$, and a uniform schedule.
    \textbf{a}, Full native parameter-gradient norm
    $\|\nabla_{\gamma,\beta}C\|$.
    \textbf{b}, Cost-parameter component $\|\nabla_\gamma C\|$.
    \textbf{c}, Mixer-parameter component $\|\nabla_\beta C\|$.
    \textbf{d}, Gradient anisotropy $A_{\gamma/\beta}$; positive values indicate cost-parameter-dominated sensitivity and negative values indicate mixer-parameter-dominated sensitivity. Contours indicate constant total physical evolution time
    $T=p(\gamma+\beta)$ for the uniform schedule.
    Panels \textbf{a--c} use logarithmic colour scales.}
    \label{fig:component-gradient-heatmap}
\end{figure*}

\subsection{Numerical setting}
\label{sec:numerics}

We test the strength--imbalance framework using exact state-vector simulations of QAOA for MaxCut, taking unweighted instances as the primary setting and introducing generic real-weighted instances as robustness and control tests.

For an unweighted graph $G=(V,E)$, MaxCut seeks a bipartition of the vertices that maximizes the number of edges crossing between the two sets, where $V$ and $E$ denote the vertex and edge sets, respectively. We use
the cost and mixer Hamiltonians
\begin{equation}
H_c=\sum_{(i,j)\in E}\frac{1-Z_iZ_j}{2},
\qquad
H_m=\sum_{i=1}^{n}X_i ,
\label{eq:results-hamiltonians}
\end{equation}
so that the eigenvalues of $H_c$ are the corresponding cut values. Hence $\|H_c\|=C_{\max}$, where $C_{\max}$ is the maximum cut value, while $\|H_m\|=n$. Circuits are initialized in $|+\rangle^{\otimes n}$, and the QAOA objective is to maximize
\[
C=\langle H_c\rangle .
\]

Unless stated otherwise, we use uniform schedules at depth $p=3$ on dense, unweighted MaxCut instances, with $n=10$ providing the representative landscapes shown in the main text. This depth captures nontrivial interference while keeping the landscape sufficiently simple to resolve its coarse dynamical organization. We then test robustness at $p=10$.

For a uniform schedule, every layer has the same $(S_\ell,\alpha_\ell)$, so each point in the dynamical plane corresponds to a single depth-$p$ QAOA circuit.

We characterize the native parameter response by
\begin{equation}
\|\nabla_{\gamma,\beta}C\|
=
\left[
\sum_{\ell=1}^{p}|\partial_{\gamma_\ell}C|^2+
\sum_{\ell=1}^{p}|\partial_{\beta_\ell}C|^2
\right]^{1/2},
\label{eq:full-gradient-norm}
\end{equation}
and compare different depths using
\begin{equation}
G_{\mathrm{RMS}}
\coloneqq
\frac{\|\nabla_{\gamma,\beta}C\|}{\sqrt{2p}},
\label{eq:gradient-rms}
\end{equation}
which normalizes for the number of native variational parameters. Further numerical details, including gradient evaluation and component-resolved
diagnostics, are given in Methods.

\begin{figure*}
    \centering
    \includegraphics[width=0.95\textwidth]
    {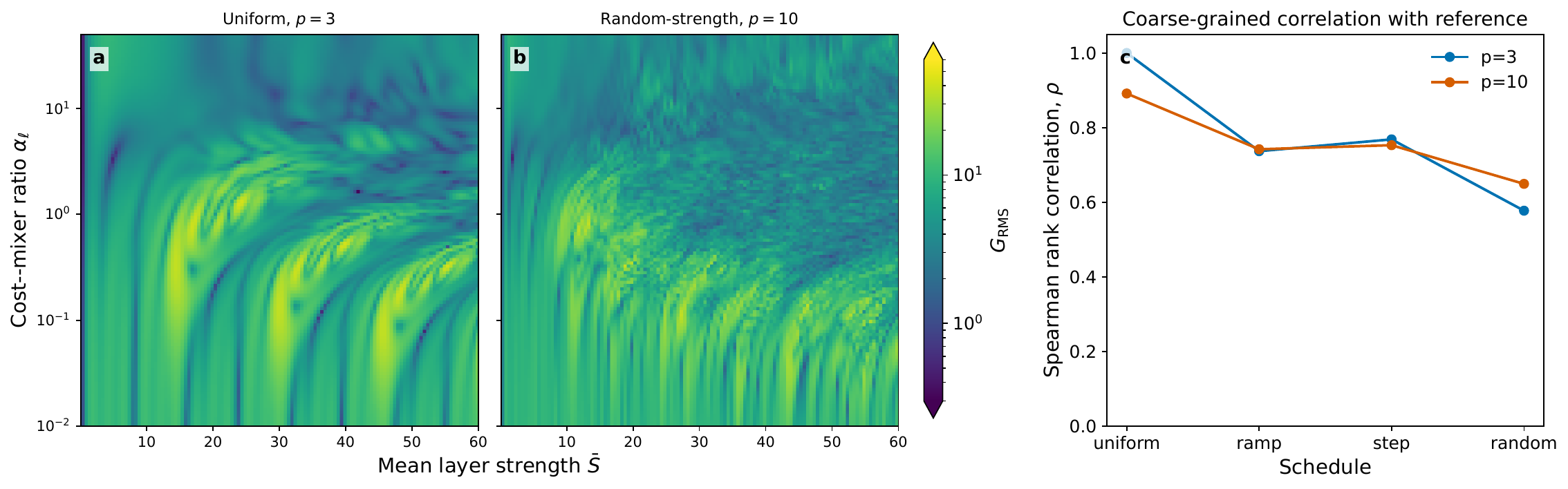}
        \caption{\justifying \textbf{Robustness of the gradient landscape to depth and
        schedule structure.}
        Results for a representative dense, unweighted MaxCut instance with $n=10$.
        \textbf{a}, Uniform schedule at $p=3$, used as the reference landscape.
        \textbf{b}, Representative random-strength schedule at $p=10$, with layerwise strengths varied using a fixed seeded random profile with $\sigma=0.4$ and normalized to preserve the mean layer strength $\bar S$, while keeping the layerwise imbalance fixed.
        \textbf{c}, Spearman rank correlation of the coarse-grained gradient landscape for each tested schedule family and depth with the $p=3$ uniform reference. The full set of schedule-resolved landscapes is shown in Supplementary Fig.~1.}
    \label{fig:schedule-depth-robustness}
\end{figure*}

\subsection{Dynamical organization of gradient response}
\label{subsec:numerics-gradient-organization}

We first ask whether the strength--imbalance representation exposes a systematic organization of the QAOA gradient landscape. 
Figure~\ref{fig:component-gradient-heatmap} reveals broad suppressed and responsive regions in the $(S_\ell,\alpha_\ell)$ plane. Moving beyond weak-drive regime produces an extended region of enhanced response, while varying $\alpha_\ell$ changes the relative sensitivity to the cost and mixer parameters. Here and below, ``suppressed'' and ``enhanced'' denote comparatively low and high response within a given landscape rather than universal absolute gradient thresholds.

The constant-time contours in Fig.~\ref{fig:component-gradient-heatmap}(a) provide a complementary view of the landscape: the same total evolution
time can correspond to noticeably different gradient responses depending on the cost--mixer allocation. Thus, circuits with the same time budget can occupy distinct dynamical regimes in the strength-imbalance plane.

These results establish strength and cost-mixer ratio as natural coordinates for the coarse organization of QAOA gradient response.

\subsection{Robustness to depth and schedule structure}
\label{subsec:numerics-schedule-depth}

We next test whether this dynamical organization persists as the circuit depth and the layerwise strength profile are varied simultaneously. To keep the visualization tractable, we consider uniform, ramp, matched-step, and random-strength schedules, with $S_\ell$ varying across layers at fixed $\alpha_\ell=\alpha$ and with the total applied strength normalized to $\sum_\ell S_\ell=p\bar S$. This introduces controlled nonuniformity while preserving a two-dimensional $(\bar S,\alpha)$
description.

As a representative case, Fig.~\ref{fig:schedule-depth-robustness} compares the $p=3$ uniform landscape with a $p=10$ random-strength schedule. Increasing the depth and introducing layerwise variation substantially reorganize the fine interference pattern, while the broad locations of suppressed and enhanced response remain recognizable.

The Spearman correlations in Fig.~\ref{fig:schedule-depth-robustness}(c) further support the similarity of the broad gradient landscape to the $p=3$ uniform reference across the tested depths and schedule families. The complete schedule-resolved comparison is provided in Supplementary Fig.~1.

In addition, a stronger test in Supplementary Sec.~I\,B allows both $S_\ell$ and $\alpha_\ell$ to vary while fixing only the global coordinates $(\bar S,\alpha_{\rm tot})$. This tests whether the same coarse organization survives after the prescribed schedule constraint is removed.

Taken together, these comparisons indicate that depth and a non-uniform schedule primarily reorganize the fine interference structure, while the broad dynamical location of responsive and suppressed regimes is more robust. 

\begin{figure*}
    \centering
    \includegraphics[width=0.80\textwidth]
    {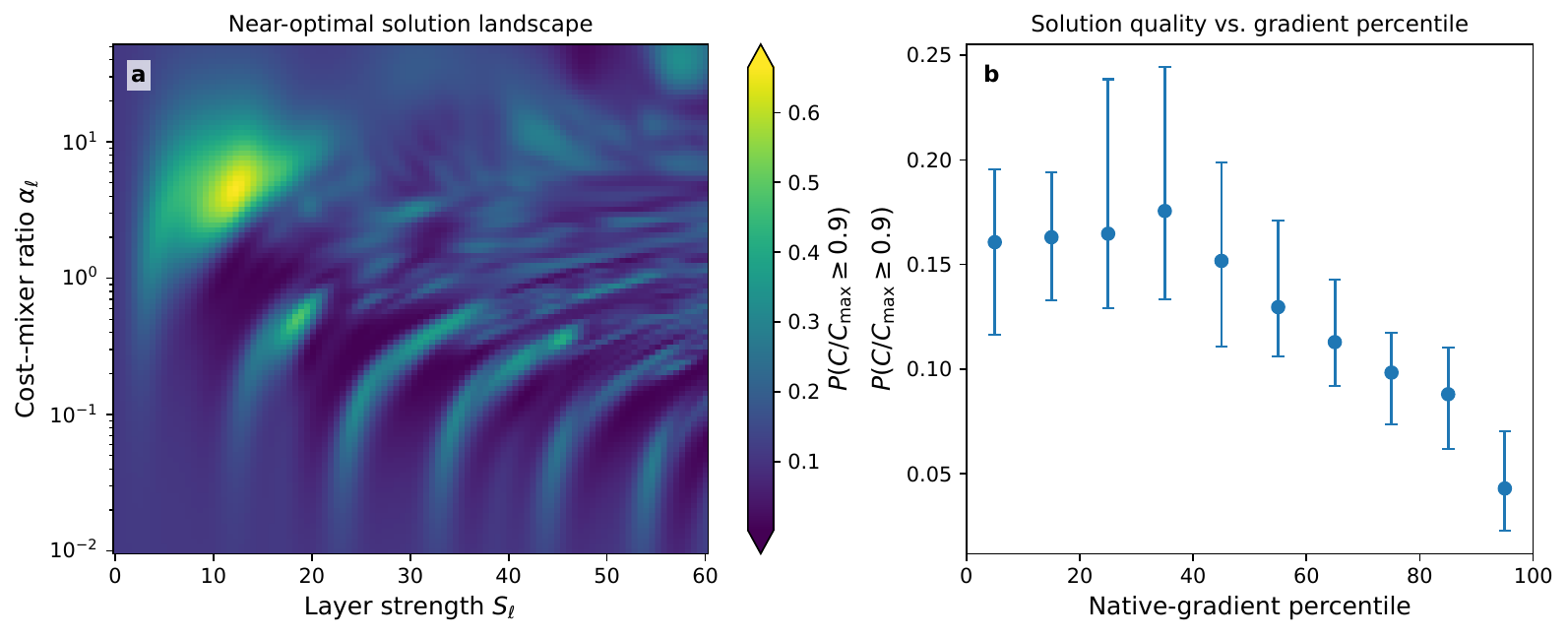}
\caption{\justifying
    \textbf{Solution quality and local gradient response.}
    Results for a representative dense, unweighted MaxCut instance with $n=10$, depth $p=3$, and a random-strength schedule at fixed layerwise imbalance.
    \textbf{a}, Near-optimal sampling probability
    $P_{\rm tail}(0.9)$ over the strength--imbalance plane.
    \textbf{b}, $P_{\rm tail}(0.9)$ as a function of native-gradient percentile across the landscape in \textbf{a}. Markers show the median within ten percentile bins and error bars the interquartile range.}
    \label{fig:solution-quality-random-p3}
\end{figure*}

\subsection{Solution quality and local gradient response}
\label{subsec:numerics-solution-quality}

Solution quality provides a complementary view of the dynamical landscape.
For a given set of QAOA parameters, measurement of the output state
$|\psi\rangle$ in the computational ($Z$) basis returns a bit string $z$
with probability $P(z)=|\langle z|\psi\rangle|^2$, corresponding to a
MaxCut solution with cut value $C(z)$. We quantify near-optimal sampling by
\[
P_{\rm tail}(0.9)
=
P\!\left(C(z)\geq0.9\,C_{\max}\right),
\]
the probability of sampling a solution whose cut value is at least $90\%$
of the optimum.

Let $\Pi_{\mathrm{tail}}$ denote the projector onto computational-basis
states satisfying $C(z)\geq0.9\,C_{\max}$, and let
$P_0=\langle\psi_0|\Pi_{\mathrm{tail}}|\psi_0\rangle$, with
$|\psi_0\rangle=|+\rangle^{\otimes n}$, denote the corresponding
near-optimal sampling probability of the initial state.

For a uniform schedule, all layers share the same values of the dynamical coordinates,
$S_\ell=S$ and $\alpha_\ell=\alpha$.
At small layer strength $S$, we expand the QAOA evolution in the
corresponding norm-weighted cost and mixer actions to second order.
The first-order contribution to $P_{\mathrm{tail}}(0.9)$ vanishes, leaving
\begin{equation}
P_{\mathrm{tail}}(0.9)-P_0
=
A_{G,p}\,
S^2\frac{\alpha}{(1+\alpha)^2}
+\mathcal{O}(S^3),
\label{eq:ptail_weak}
\end{equation}
where, for unweighted MaxCut,
\begin{equation}
A_{G,p}
=
\frac{4p(p+1)}{nC_{\max}}
P_0
\left(
\mu_{\mathrm{tail}}-\frac{|E|}{2}
\right),
\label{eq:Agp}
\end{equation}
with
\[
\mu_{\mathrm{tail}}
:=
\mathbb{E}\!\left[
C(z)\,\middle|\,C(z)\geq0.9\,C_{\max}
\right].
\]
This leading quadratic behavior is consistent with the general
small-angle QAOA probability expansion of Hadfield
\textit{et al.}~\cite{Hadfield2023}.

The factor $S^2\alpha/(1+\alpha)^2$ has the same strength--imbalance
dependence as the leading mixed cost--mixer contribution identified in the
Magnus analysis of Sec.~\ref{subsec:imbalance-bch-structure}.

For each fixed $\alpha$ and target probability $P_\star>P_0$, we define the first favorable region as the first contiguous interval in strength, encountered as $S$ increases from zero, for which
\[
P_{\mathrm{tail}}(0.9)\geq P_\star.
\]
We denote its lower and upper boundaries by
\[
S_-(\alpha;P_\star)\leq S\leq S_+(\alpha;P_\star).
\]
This convention distinguishes the first low-strength favorable region from possible disconnected regions appearing at stronger drive.

Parameterizing the target probability as
\[
P_\star=P_0+\eta(1-P_0),
\]
where \(\eta\in(0,1)\) sets how much the target probability is increased above $P_0$, and solving Eq.~\eqref{eq:ptail_weak} to leading order gives the perturbative prediction for the lower boundary
\begin{equation}
S_-(\alpha;\eta)
\simeq
\frac{1+\alpha}{\sqrt{\alpha}}
\sqrt{
\frac{\eta(1-P_0)}
{A_{G,p}}
}.
\label{eq:Sminus}
\end{equation}
Using the explicit form of $A_{G,p}$, this can be written as
\begin{equation}
\begin{aligned}
S_-(\alpha;\eta)
&\simeq
\underbrace{\frac{1+\alpha}{\sqrt{\alpha}}}_{\text{imbalance}}
\underbrace{\frac{1}{\sqrt{p(p+1)}}}_{\text{depth}}
\sqrt{\eta}\,
\underbrace{
\sqrt{
\frac{nC_{\max}(1-P_0)}
{4P_0\left(\mu_{\mathrm{tail}}-|E|/2\right)}
}
}_{\text{instance spectrum}} .
\end{aligned}
\label{eq:Sminus_factorized}
\end{equation}

Thus, the perturbative boundary is minimized at balanced driving, $\alpha=1$, and shifts to larger strength as the dynamics become increasingly c. The expansion determines only this weak-drive onset; the upper boundary $S_+$ can depend on higher-order dynamics, schedule-dependent interference, and spectral recurrence. The dependence
of $S_-$ on $\alpha$ and circuit depth is explicit, while its absolute scale remains instance dependent through $C_{\max}$, $P_0$, and $\mu_{\mathrm{tail}}$. The result therefore provides a structural prediction rather than an instance-independent parameter-selection rule.

Expressed in terms of total evolution time, for fixed \(\alpha\) and problem instance one has \(T\propto pS\). Hence, \(S_-\propto[p(p+1)]^{-1/2}\) implies \(T_-\propto\sqrt{p/(p+1)}\). Thus, the perturbatively predicted onset approaches a depth-independent total evolution time as \(p\) increases.

A derivation of Eqs.~\eqref{eq:ptail_weak}--\eqref{eq:Sminus} is given in Methods, Sec.~\ref{subsec:perturbative-onset}. We validate the
second-order prediction numerically across independent dense and sparse MaxCut instances, unweighted and real-weighted cost Hamiltonians, several circuit depths, and both uniform and random-strength schedules. For unweighted uniform schedules, the numerical onset is compared directly with Eq.~\eqref{eq:Sminus}, whereas random-strength schedules and real-weighted instances are tested against the corresponding generalized second-order prediction. Agreement is strongest while the predicted onset remains within the perturbative small-\(S\) regime, with deviations increasing as the onset shifts to larger strength.


Eq.~\eqref{eq:Sminus} provides an analytical characterization of the perturbative onset, but its direct evaluation requires problem-specific quantities that are generally not known a priori, including \(C_{\max}\) and the conditional mean \(\mu_{\mathrm{tail}}\) of the corresponding high-cost sector. In practice, this requirement can be addressed in several ways. One option is to approximate \(C_{\max}\) using a classical MaxCut algorithm such as Goemans--Williamson. Alternatively, these quantities can be replaced by statistics obtained from an upper quantile of the random-cut distribution, estimated directly from uniformly sampled bit strings. In this formulation, the leading-order onset can be expressed in terms of sampled tail statistics without requiring explicit knowledge of \(P_0\) or \(C_{\max}\). Details of the construction are given in Methods, Sec.~\ref{subsec:perturbative-onset}, with numerical validation reported in Supplementary Sec.~I\,D.

Figure~\ref{fig:solution-quality-random-p3} shows the solution-quality landscape for a random-strength schedule. A localized near-optimal solution region appears at intermediate-to-cost-biased imbalance, indicating that favorable solutions occupy a restricted portion of the strength--imbalance plane. This near-optimal region closely matches the
dynamical expectation summarized schematically in
Fig.~\ref{fig:regimes}.

The gradient-percentile comparison provides a complementary view: the largest typical $P_{\rm tail}(0.9)$ occurs at intermediate gradient response, while lower values are found in the highest-gradient
percentiles.

The same qualitative organization persists across uniform and random-strength schedules, increased depth, and real-weighted cost Hamiltonians, as shown in Supplementary Sec.~I\,C. A stronger test with fully nonuniform schedules shows that near-optimal circuits remain statistically concentrated in approximately the same coarse region of \((\bar S,\alpha_{\rm tot})\), despite substantial schedule-to-schedule variation within that region (Supplementary Sec.~I\,G).

At fixed \((\bar S,\alpha_{\rm tot})\), schedules share the same global cost and mixer actions. Their differing solution quality therefore reflects the layerwise distribution of this action, indicating that the global coordinates locate the coarse favorable regime while detailed scheduling modulates performance within it.

\subsection{Solution-bearing regions and native-coordinate accessibility}
\label{subsec:numerics-solution-preimage}

\begin{figure*}
    \centering
    \includegraphics[width=\linewidth]
    {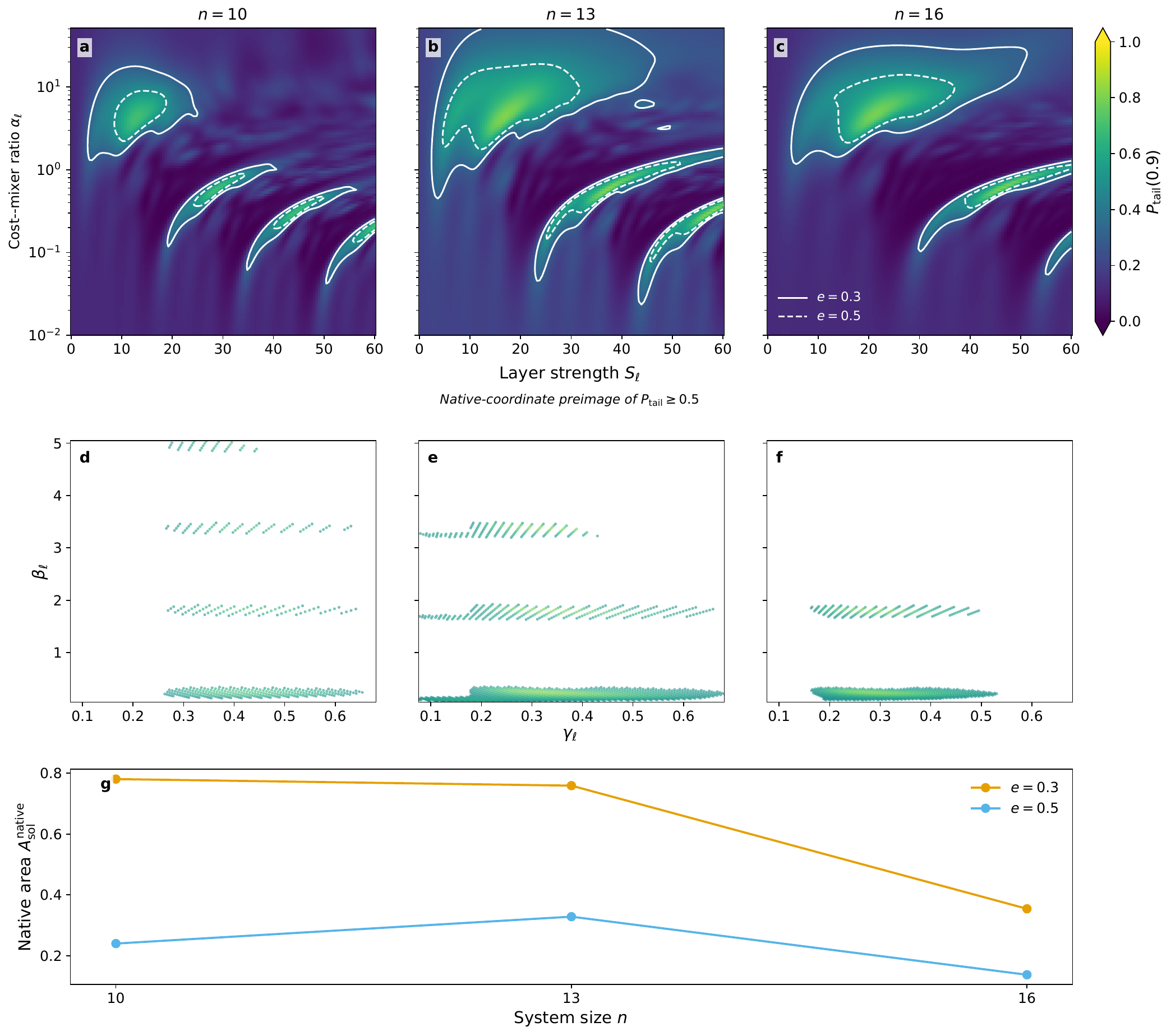}
    \caption{\justifying \textbf{Near-optimal regions and their native-coordinate
    preimages.}
    \textbf{a--c}, $P_{\rm tail}(0.9)$ over the strength--imbalance plane for representative dense, unweighted MaxCut instances with
    $n=10$, $13$, and $16$, using $p=3$ uniform schedules and a common colour scale. Contours mark
    $P_{\rm tail}(0.9)=0.3$ and $0.5$.
    \textbf{d--f}, Native $(\gamma_\ell,\beta_\ell)$ preimages of the $P_{\rm tail}(0.9)\geq0.5$ regions, shown with common axis limits.
    \textbf{g}, Jacobian-integrated native area
    $A_{\rm sol}^{\rm native}$ for the regions satisfying $P_{\rm tail}(0.9)\geq0.3$ and $0.5$.}
    \label{fig:solution-preimage}
\end{figure*}

Figure~\ref{fig:solution-preimage}(a--c) shows \(P_{\rm tail}(0.9)\) for representative dense, unweighted instances with \(n=10\), \(13\), and \(16\). Regions satisfying fixed \(P_{\rm tail}(0.9)\) thresholds remain extended in the strength--imbalance plane, while shifting and deforming toward larger layer strengths as the system size increases.
In the native \((\gamma_\ell,\beta_\ell)\) coordinates, however, the preimages of the \(P_{\rm tail}(0.9)\geq0.5\) regions are more compressed at \(n=16\) than at \(n=10\) (d--f). The Jacobian-integrated native area shows the same overall contraction for both quality thresholds, although the finite-size variation is not strictly monotonic (g).
These results provide a numerical realization of the geometric contraction predicted in Sec.~\ref{subsec:size-scaling}: over the sizes studied, a favorable region can remain extended and drift in dynamical coordinates while occupying a smaller area in the native parameterization. An analogous contraction for high-response gradient regions across dense and sparse graph families is shown in Supplementary Sec.~I\,H.

\subsection{Strong-drive dynamics and cost-spectrum recurrence}
\label{subsec:numerics-strong-drive}

\begin{figure*}
    \centering
    \includegraphics[width=\linewidth]
    {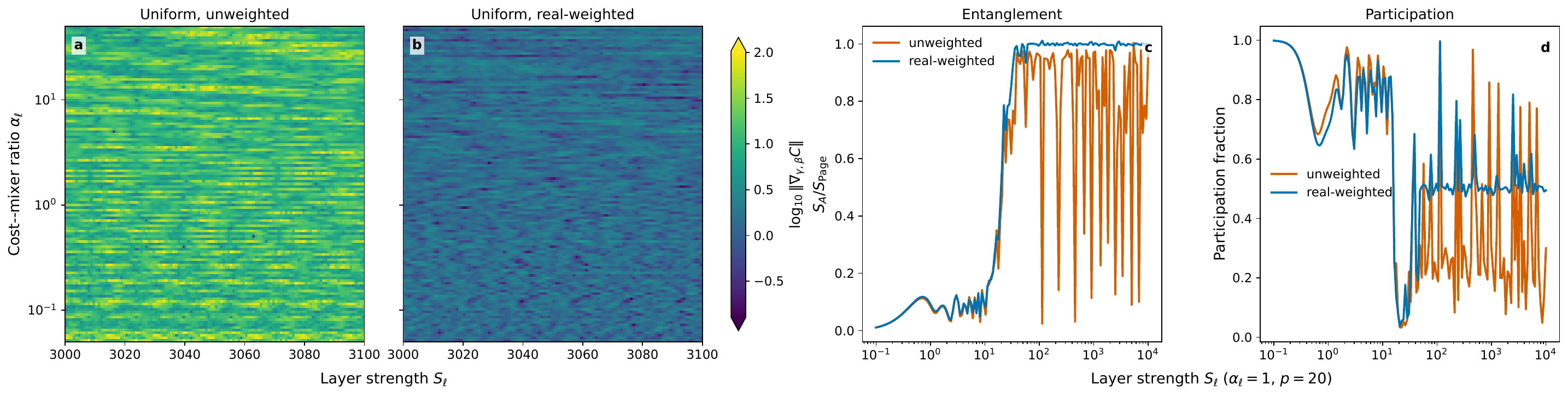}
    \caption{\justifying \textbf{Strong-drive dynamics and the role of cost-spectrum
    recurrence.}
    \textbf{a,b}, Native-gradient response over the same far strong-drive window for a representative dense MaxCut topology with $n=10$ and $p=3$:
    \textbf{a}, unweighted MaxCut; \textbf{b}, the same topology with generic real edge weights. The panels share a common colour scale.
    \textbf{c}, Half-system entanglement entropy normalized by the Page value, and \textbf{d}, normalized computational-basis participation fraction for representative unweighted and real-weighted instances at $\alpha_\ell=1$ and $p=20$. Additional depth, size, and weight-realization controls are
    provided in the Supplementary Information.}
    \label{fig:strong_drive}
\end{figure*}

At strong drive, repeated winding of the QAOA phases creates increasingly complex interference structure. In unweighted MaxCut, the integer-valued cost spectrum makes the cost evolution exactly $2\pi$-periodic in the native cost parameter~\cite{Sureshbabu2024WeightedQAOA}. This periodicity is distorted by the nonlinear mapping to strength--imbalance coordinates, as discussed in Supplementary Sec.~I\,I. To probe its role, we compare unweighted MaxCut with a generic real-weighted version of the same graph topology, for which no common \(2\pi\) cost-phase recurrence is present~\cite{ShaydulinWeightedMaxCut2023}. Figures~\ref{fig:strong_drive}(a),(b) show that, far into the strong-drive regime, the unweighted landscape retains pronounced structure and a comparatively stronger gradient response, whereas the real-weighted landscape is more homogeneous and suppressed. Recurrence-related structure is resolved more clearly over a lower-strength window in the Supplementary Information.

The state diagnostics show a consistent distinction. At \(\alpha_\ell=1\), both models develop near-Page entanglement at sufficiently large strength (Fig.~\ref{fig:strong_drive}c,d). The real-weighted evolution then remains broadly distributed, with participation near \(1/2\), whereas the unweighted evolution exhibits repeated entropy collapses and participation drops, consistent with partial relocalization associated with the exact recurrence of the cost evolution.

The persistent near-Page entanglement and broad participation of the real-weighted evolution are consistent with ergodic-like state spreading and resemble behavior associated with Floquet heating in periodically driven many-body systems. Establishing ergodicity or thermalization in finite-depth QAOA, however, would require additional diagnostics.

To determine whether this state-spreading behavior extends beyond the balanced line \(\alpha_\ell=1\), we examine the strength required to reach fixed fractions of the Page entropy across the full strength--imbalance plane. Across dense real-weighted instances, the strength required to reach fixed fractions of the Page entropy generally decreases as the dynamics become more cost dominated, although the crossover becomes less regular in the strongly mixer-dominated regime. We interpret these thresholds as crossover boundaries rather than sharp transitions; the full construction and system-size dependence are shown in Supplementary Sec.~I\,K. This trend is consistent with the distinct roles of the two generators: the transverse-field mixer factorizes into local unitaries across any bipartition and therefore leaves bipartite entanglement unchanged, whereas cost interactions crossing the bipartition can generate it.

\section{Discussion}

Our results show that QAOA gradient response can be organized dynamically through the layer strength $S_\ell$ and the cost--mixer ratio $\alpha_\ell$. These variables separate the overall scale of the applied evolution from the relative contribution of the two noncommuting generators. Within this representation, weak gradients arise through distinct mechanisms: insufficient evolution at weak drive, reduced cost--mixer competition under strong single-generator dominance, and more complex higher-order dynamics at strong drive. The trace-speed analysis further separates the capacity for parameter-induced state motion from the gradient response actually realized in the cost expectation.

The numerical results indicate that this organization is robust at a coarse level across the depths and schedule constructions considered, while finer interference structure remains schedule dependent. Near-optimal samples are concentrated primarily at intermediate strength in a balanced-to-cost-biased region, and their relation to gradient magnitude is nonmonotonic. The factorized onset expression makes the predictive scope explicit:
the dependence of the lower boundary on cost--mixer imbalance and circuit depth is fixed analytically, while its absolute scale remains instance dependent through $C_{\max}$, $P_0$, and $\mu_{\mathrm{tail}}$.
For the small systems studied here, these quantities are obtained exactly. The onset prediction therefore captures how the weak-drive boundary varies with the dynamical coordinates and depth, conditional on
instance-specific spectral information, rather than providing a standalone instance-independent parameter-selection rule. At the same time, the native-parameter preimages of responsive regions contract as the generator norms grow with system size. This identifies an accessibility effect in native coordinates, but does not by itself establish asymptotic gradient suppression.

For unweighted MaxCut, the commensurate cost spectrum produces recurrent structure and partial relocalization, whereas generic real weights remove a common cost recurrence and lead to substantially broader state spreading. The accompanying entanglement and participation behavior is suggestive of mixing and thermalization-like dynamics, indicating that the emergence of this behavior depends on spectral and circuit structure rather than on strong drive alone. A more direct connection between these dynamical regimes and established barren-plateau scaling therefore remains to be determined.

The present numerical study is restricted to small-system MaxCut instances and does not establish whether the same regime structure persists at substantially larger sizes or for other classes of problem Hamiltonians. Our analysis characterizes the underlying QAOA landscape independently of any particular classical optimization algorithm; how different optimizers exploit or resolve these dynamical regimes remains an important direction for future work. This is especially relevant for broader schedule distributions, where schedule-dependent structure may influence problem-aware or adaptive
optimization in native parameter space. A further question is whether these regimes remain accessible under realistic control constraints and hardware noise. More broadly, extending the strength--imbalance framework across problem classes, schedule families, and system sizes may help distinguish universal features of QAOA trainability from problem-dependent behavior.

\section{Methods}
\label{sec:methods}

\subsection{QAOA as Controlled Quantum Dynamics}
\label{sec:formulation}

The Quantum Approximate Optimization Algorithm  (QAOA) \cite{Farhi2014} is a hybrid quantum-classical algorithm designed to find approximate solutions to combinatorial optimization problems by preparing a parametrized quantum state through alternating applications of a problem-dependent- and a so-called mixing Hamiltonian. Its parameters are then variationally optimized on a classical computer to maximize the expectation value of the objective function.
We formulate QAOA within the framework of controlled quantum dynamics, where the circuit is viewed as a time-dependent evolution generated by a pair of controllable Hamiltonians. This perspective makes explicit that QAOA is fundamentally a \emph{finite-time control process}, and that its behavior is governed by how the available evolution time is allocated across layers.
In this setting, the objective is encoded in a cost Hamiltonian \(H_c\), while the mixer Hamiltonian \(H_m\) generates transitions between computational-basis states. Optimizing the variational schedule aims to prepare a state with a favorable expectation value of \(H_c\), corresponding to low energy for minimization problems or high energy for maximization problems.

We assume that a quantum system evolves under a time-dependent Hamiltonian of the form
\begin{equation}
H(t) = H_0 + \sum_{k \in \{c,m\}} u_k(t)\, H_k .
\label{eq:time_dependent_H}
\end{equation}
Here $H_c$ and $H_m$ denote the cost and mixer Hamiltonians, respectively. 
For clarity, we neglect the drift term $H_0$. 
The controlled Hamiltonian therefore reduces to
\begin{equation}
H(t) = u_c(t)\,H_c + u_m(t)\,H_m .
\label{eq:control_hamiltonian}
\end{equation}
We assume that the overall energy scale associated with the cost Hamiltonian $H_c$ is fixed, so the control coefficients, e.g.~$u_c(t)$, are not absorbed into the Hamiltonians.
The associated dynamical Lie algebra
\begin{equation}
\mathfrak{g} := \mathrm{Lie}\{ iH_c,\, iH_m \}
\label{eq:lie_algebra}
\end{equation}
characterizes the directions generated by the available controls. Under standard controllability conditions, it determines the connected set of unitaries that can be reached in principle. Related geometric work on quantum neural networks has similarly emphasized that reachability alone is insufficient to characterize learnability~\cite{Ngairangbam2026Reachability}. For a finite control horizon, however, which parts of this reachable set can be accessed also depends on the available evolution time.

As we emphasize below, what can be reached in practice is constrained not only by algebraic structure, but also by the \emph{available evolution time}.

We adopt a time-based parametrization in which control amplitudes are absorbed into effective gate durations.
To this end, we restrict ourselves to the simplest case of a piecewise-constant control scheme.
Under this convention, the QAOA variational parameters are interpreted directly as nonnegative evolution times. In the absence of amplitude constraints, this reparametrization does not alter the reachable unitary set, but makes explicit that the total physical evolution time is the primary resource governing the dynamics.

We fix a total evolution time $T$ over the interval $[0,T]$ and partition it into $p$ subintervals~\cite{CamposVenuti2021OptimalControl, Brady2021OptimalProtocols},
\begin{equation}
[0,T] = \bigcup_{\ell=1}^{p} [t_{\ell-1}, t_\ell],
\end{equation}
with layer durations
\begin{equation}
\tau_\ell := t_\ell - t_{\ell-1}.
\end{equation}
On layer $\ell$, the duration $\tau_\ell$ is divided between cost and mixer evolutions. We identify the QAOA parameters with these durations,
\[
\gamma_\ell > 0,
\qquad
\beta_\ell > 0,
\]
where $\gamma_\ell$ and $\beta_\ell$ are the evolution times under cost- and mixer Hamiltonian, respectively,
so that
\begin{equation}
\tau_\ell = \gamma_\ell + \beta_\ell.
\label{eq:time_partition}
\end{equation}
The total evolution time is therefore
\begin{equation}
T = \sum_{\ell=1}^{p} \tau_\ell,
\label{eq:total_time}
\end{equation}
and the resulting depth-$p$ unitary takes the standard product form
\begin{equation*}
U_p
=
\prod_{\ell=1}^{p}
e^{-i \beta_\ell H_m}
e^{-i \gamma_\ell H_c}.
\end{equation*}

In this formulation, the circuit depth \( p \) determines how the total evolution time \( T \) is distributed and ordered across layers. While the reachable set is determined by the total evolution time, the ordering and distribution of this time influence the effective dynamics. In particular, depth controls how noncommutative contributions accumulate through time-ordered evolution, shaping the structure of the dynamics even at fixed total time~\cite{CamposVenuti2021OptimalControl,Brady2021OptimalProtocols}.



\subsection{Geometric Origin of Gradient Behaviour}
\label{sec:structure}

The geometric displacement of a quantum state under the applied dynamics can be quantified using the trace distance. While the BCH/Magnus expansion characterizes the effective generator and the corresponding directions of motion~\cite{Magnus1954,Blanes2009}, geometric and quantum-speed-limit bounds constrain how far the state can move in finite time~\cite{AnandanAharonov1990,Taddei2013,DeffnerCampbell2017}. We use these bounds to relate finite-time state displacement to parameter sensitivity in QAOA.



\paragraph{Trace-distance speed bound.}
The geometric displacement of a quantum state is bounded by its dynamical response under the applied Hamiltonian. Let $\rho(t)$ denote the evolving density operator. The trace distance between two states is
\begin{equation}
D(\rho,\sigma) = \tfrac12\|\rho-\sigma\|_1 .
\label{eq:trace-distance}
\end{equation}
If $\rho(t)$ evolves under a bounded Hamiltonian $H(t)$, then \cite{NielsenChuang2000,DeffnerCampbell2017} 
\begin{equation}
D(\rho(t),\rho(0))
\le
\int_0^t v_{H(s)}(\rho(s))\, \mathrm{d}s,
\label{eq:trace-speed}
\end{equation}
where the trace-speed is defined as~\cite{GessnerSmerzi2018}
\begin{equation}
v_{H(t)}(\rho(t)) := \tfrac12\|[H(t),\rho(t)]\|_1.
\label{eq:trace-speed-definition}
\end{equation}
This is consistent with geometric quantum-speed-limit formulations~\cite{AnandanAharonov1990,Taddei2013,DeffnerCampbell2017}.
For pure states $\rho_\psi=\ket{\psi}\!\!\bra{\psi}$, we have
\begin{equation}
\tfrac12\|[H,\rho_\psi]\|_1 = \Delta_\psi H \le \|H\|,
\end{equation}
where
\begin{equation}
\Delta_\psi H
=
\sqrt{
\langle \psi|H^2|\psi\rangle
-
\langle \psi|H|\psi\rangle^2
}
\end{equation}
is the standard deviation of $H$ in the state $\ket{\psi}$.
For arbitrary density operators, the bound $v_H(\rho)\le \|H\|$ gives
\begin{equation}
D(\rho(t),\rho(0)) \le t\,\|H\|.
\label{eq:qsl-displacement}
\end{equation}
Thus, the geometric displacement is bounded by the accumulated dynamical response, consistent with quantum speed-limit bounds on finite-time evolution~\cite{MandelstamTamm1945,AnandanAharonov1990,DeffnerCampbell2017}.

\paragraph{Layer-wise displacement bound.}
The general trace-distance bound can be applied directly to a single QAOA layer. Let $\rho_{\ell-1}$ denote the state before layer $\ell$, $\tilde{\rho}_{\ell}$ the intermediate state after the cost evolution, and $\rho_\ell$ the state after the subsequent mixer evolution. Using the triangle inequality together with the bound above,
\begin{equation}
\begin{aligned}
D_\ell
&:=D(\rho_{\ell-1},\rho_\ell) \\
&\le
D(\rho_{\ell-1},\tilde{\rho}_{\ell})
+
D(\tilde{\rho}_{\ell},\rho_\ell) \\
&\le
\gamma_\ell\,\|H_c\|
+
\beta_\ell\,\|H_m\|.
\end{aligned}
\label{eq:layer-trace-bound}
\end{equation}

Using the Lipschitz continuity of the objective function defined in Eq.~\eqref{eq:objective} with respect to the trace distance, we obtain
\begin{equation}
|C(\rho_{\ell})-C(\rho_{\ell-1})|
\le
2\|H_c\|\,D_\ell.
\label{eq:cost-change-bound}
\end{equation}
This bound relates the local change in the objective to the state displacement during that layer. The layer-wise trace-distance bound identifies
$\gamma_\ell\|H_c\|$ and $\beta_\ell\|H_m\|$
as natural norm-weighted cost and mixer actions, providing a geometric motivation for the strength--imbalance variables used in the main text.



\paragraph{Gradient bound from dynamical response.}
The gradient of the objective is determined by how the variational state responds to changes in the circuit parameters. Let 
\begin{equation}
\rho_\theta = U(\theta)\rho_0 U^\dagger(\theta)
\end{equation}
be the parameterized quantum state prepared by the ansatz, where $U(\theta)$ is a parameterized circuit applied to a fixed initial state $\rho_0$. 

The objective function is then given by
\begin{equation}
C(\theta)=\mathrm{Tr}(\rho_\theta H_c),
\label{eq:objective}
\end{equation}
which is the expectation value of the cost Hamiltonian $H_c$. For the MaxCut convention used here, maximizing $C(\theta)$ favors states with larger cut values. The parameter derivative of the state can be written as
\begin{equation}
\frac{\mathrm{d}\rho_\theta}{\mathrm{d}\theta} = -i[G_\theta,\rho_\theta],
\end{equation}

where
\begin{equation}
G_\theta
=
i(\partial_\theta U)U^\dagger
\label{eq:generator-parameter-variation}
\end{equation}
is the Hermitian generator associated with variations in the parameter
$\theta$.
For a parameterized gate generated by $H_\theta$, we write the circuit as
\[
U
=
U_{>\theta}\,e^{-i\theta H_\theta}\,U_{<\theta},
\]
where $U_{>\theta}$ and $U_{<\theta}$ denote the ordered products of gates
applied after and before the parameterized gate, respectively. It follows that
\begin{equation}
G_\theta
=
U_{>\theta}H_\theta U_{>\theta}^{\dagger}.
\end{equation}

Using the layer unitary defined in Eq.~\eqref{eq:layer-unitary}, with
$U=U_p\cdots U_1$, the circuit around layer $\ell$ can be written as
\[
U
=
U_p\cdots U_{\ell+1}
e^{-i\beta_\ell H_m}
e^{-i\gamma_\ell H_c}
U_{\ell-1}\cdots U_1 .
\]
Hence,
\[
U_{>\gamma_\ell}
=
U_p\cdots U_{\ell+1}e^{-i\beta_\ell H_m},
\qquad
U_{>\beta_\ell}
=
U_p\cdots U_{\ell+1}.
\]
The corresponding QAOA generators are therefore
\begin{equation}
G_{\gamma_\ell}
=
U_{>\gamma_\ell}H_cU_{>\gamma_\ell}^{\dagger},
\qquad
G_{\beta_\ell}
=
U_{>\beta_\ell}H_mU_{>\beta_\ell}^{\dagger}.
\label{eq:generators}
\end{equation} As in Sec.~\ref{subsec:bch-depth}, this segment can be represented by a Magnus generator whose commutator structure dresses the corresponding cost or mixer generator and thereby enters the gradient response.
The gradient is then
\begin{equation}
\partial_\theta C
=
i\,\mathrm{Tr}\!\big(\rho_\theta [G_\theta,H_c]\big).
\label{eq:grad-commutator}
\end{equation}
Using Hölder's inequality, we finally have
\begin{equation}
|\partial_\theta C|
\le
2\|H_c\|\,v_{G_\theta}(\rho_\theta).
\label{eq:gradient-trace-bound}
\end{equation}
This bound shows that gradients are limited by the dynamical response of the state to the generator $G_\theta$. 
This viewpoint is closely related to observable-speed-limit bounds, which constrain the rate of change of expectation values under quantum dynamics~\cite{GarciaPintos2022,MohanPati2022}, and to recent work on gradient bounds and trainability diagnostics for parameterized quantum circuits~\cite{Letcher2024,Larocca2022}. In the present work, we use this geometric response bound as a diagnostic of QAOA gradient response.



\subsection{Simulation and gradient evaluation}
\label{subsec:methods-simulation}

All calculations use exact noiseless state-vector evolution, without finite-shot sampling or hardware noise. Native-parameter derivatives $\partial_{\gamma_\ell}C$ and $\partial_{\beta_\ell}C$ are evaluated by adjoint differentiation, using a forward propagation followed by a reverse sweep through the circuit to obtain all $2p$ gradient components. Selected calculations were independently verified using centred finite differences, with agreement to numerical precision.

\subsection{Graph instances and Hamiltonians}
\label{subsec:methods-graphs}

For a graph $G=(V,E)$, MaxCut seeks a bipartition of the vertices that maximizes the number of edges crossing between the two sets. In the QAOA
encoding, each computational-basis bit string represents a candidate partition, with cost given by its cut value. The corresponding cost and mixer
Hamiltonians are
\[
H_c=\sum_{(i,j)\in E}\frac{1-Z_iZ_j}{2},
\qquad
H_m=\sum_{i=1}^{n}X_i .
\]
With this convention, the eigenvalues of $H_c$ are the classical cut values, so that $\|H_c\|=C_{\max}$, while $\|H_m\|=n$. The maximum cut value is obtained by exact enumeration of all computational-basis states.
 The same enumeration is used to determine the initial tail probability $P_0$ and the conditional tail mean $\mu_{\mathrm{tail}}$ entering the perturbative onset
prediction.

Representative two-dimensional landscapes are shown for a dense, unweighted $n=10$ instance. System-size and robustness analyses additionally use dense and sparse graph families. Dense graphs are constructed by removing \(n-4\) edges from the complete graph, while sparse graphs are chosen to be connected and to contain \(2n\) edges. Unless stated otherwise, the QAOA state is initialized as $|+\rangle^{\otimes n}$.

\subsection{Schedule constructions and depth robustness}
\label{subsec:methods-schedules}

For the schedule constructions used below, it is convenient to write the norm-weighted cost and mixer actions as
\begin{equation}
x_\ell:=\gamma_\ell\|H_c\|,
\qquad
y_\ell:=\beta_\ell\|H_m\|,
\end{equation}
so that
\begin{equation}
S_\ell=x_\ell+y_\ell,
\qquad
\alpha_\ell=\frac{x_\ell}{y_\ell}.
\end{equation}

Uniform schedules use the same $(S_\ell,\alpha_\ell)$ in every layer. We use $S$ for this common layer strength and $\bar S$ for the mean layer strength of a nonuniform schedule. To introduce controlled layerwise variation while retaining a two-dimensional parameter scan, the random-strength schedules vary $S_\ell$ while keeping $\alpha_\ell=\alpha$ fixed. For the schedule-robustness comparisons, the nonuniform profiles are normalized to
\begin{equation}
\sum_{\ell=1}^{p}S_\ell=p\bar S,
\end{equation}
so that schedules with different shapes are compared at the same mean layer strength.

We consider three controlled layer-strength profiles: a ramp, in which the strength increases linearly across the layers; a matched-step profile, in which successive layers differ by a constant additive increment chosen to match the total variation of the ramp; and a random-strength profile, in which the layer strengths receive fixed seeded multiplicative fluctuations.
Before normalization, these profiles are
\begin{equation}
\widetilde S_\ell^{\rm ramp}
=
\bar S
\left(
1+\delta_{\rm ramp}\frac{\ell-1}{p-1}
\right).
\end{equation}

\begin{equation}
\widetilde S_\ell^{\rm step}
=
\bar S+\delta_{\rm step}(\ell-1).
\end{equation}

\begin{equation}
\widetilde S_\ell^{\rm rand}
=
\bar S\exp(\sigma\xi_\ell).
\end{equation}
where $\ell=1,\ldots,p$ and $\xi_\ell$ is a fixed seeded pattern with zero mean and unit variance. For compactness, we write \(\widetilde S_\ell\) in the following equation to denote whichever of the three unnormalized profiles is being considered, i.e. \(\widetilde S_\ell^{\rm ramp}\), \(\widetilde S_\ell^{\rm step}\), or \(\widetilde S_\ell^{\rm rand}\). Each such profile is then rescaled to obtain the final layer strengths \(S_\ell\) with mean \(\bar S\): 
\begin{equation}
S_\ell
=
\frac{p\bar S\,\widetilde S_\ell}
{\sum_j\widetilde S_j}.
\end{equation}

The ramp uses $\delta_{\rm ramp}=0.4$, while the random-strength schedule uses $\sigma=0.4$. Because the ramp perturbation is fractional whereas the step perturbation is additive, using the same numerical variatin parameter would produce perturbations of very different magnitude. We therefore choose the step increment so that its total layer-to-layer swing matches that of the ramp at the reference strength \(S_{\rm ref}=30\):
$S_{\rm ref}=30$,
\begin{equation}
\delta_{\rm step}(p)
=
\frac{\delta_{\rm ramp}S_{\rm ref}}{p-1}.
\end{equation}
This gives comparable ramp and step perturbations while allowing the same total variation to be distributed over different circuit depths.

 Schedule robustness is evaluated at \(p=3\) and \(p=10\), as summarized in Fig.~\ref{fig:schedule-depth-robustness}. To compare the broad organization of the gradient landscapes, we use \(G_{\rm RMS}\) as defined in Eq.~\eqref{eq:gradient-rms} and average \(\log_{10}G_{\rm RMS}\) over non-overlapping \(4\times4\) blocks. The resulting landscapes are compared with the \(p=3\) uniform reference using Spearman rank correlation. This coarse graining is used only for the correlation analysis; the gradient landscapes shown in the figures are not coarse grained. The complete comparison across uniform, ramp, matched-step, and random-strength schedules is provided in Supplementary Fig.~1.

\subsection{Fully random layerwise schedules}
\label{subsec:methods-fully-random}

Fully random schedules are sampled at fixed mean layer strength $\bar S$
and fixed global cost--mixer ratio
\begin{equation}
\alpha_{\rm tot}
:=
\frac{\sum_{\ell=1}^p x_\ell}
     {\sum_{\ell=1}^p y_\ell}.
\end{equation}
The total norm-weighted cost and mixer actions are allocated across layers using two independent symmetric Dirichlet draws
\begin{equation}
\mathbf u,\mathbf v
\sim
\mathrm{Dirichlet}(\kappa,\ldots,\kappa),
\qquad
\sum_{\ell=1}^{p}u_\ell
=
\sum_{\ell=1}^{p}v_\ell
=
1 .
\end{equation}
We use $\kappa=1$, corresponding to uniform sampling over the simplex, so that no particular layer-allocation profile is favored. The components $u_\ell$ and $v_\ell$ specify the fractions of the total cost and mixer actions assigned to layer $\ell$. 

Using the norm-weighted layer actions introduced in
Sec.~\ref{subsec:methods-schedules},
\begin{equation}
x_\ell
=
\frac{\alpha_{\rm tot}}{1+\alpha_{\rm tot}}\,
p\bar S\,u_\ell,
\qquad
y_\ell
=
\frac{1}{1+\alpha_{\rm tot}}\,
p\bar S\,v_\ell .
\end{equation}

This construction allows both $S_\ell$ and $\alpha_\ell$ to vary from layer to layer while preserving the global coordinates $(\bar S,\alpha_{\rm tot})$ exactly.

For each depth, the same Dirichlet realizations are reused across the full $(\bar S,\alpha_{\rm tot})$ grid, so that changes across the plane are evaluated using the same set of layerwise allocations. The ensembles contain 88 schedules at $p=3$ and 60 schedules at $p=10$. The resulting ensemble statistics are shown in Supplementary Fig.~2.

\subsection{Solution-quality analysis}
\label{subsec:methods-solution-quality}

Solution quality is quantified by the probability of sampling a bit string whose cut value is at least $90\%$ of the optimum,
\begin{equation}
P_{\rm tail}(0.9)
\coloneqq
\sum_{z:\,C(z)/C_{\max}\geq0.9}
|\langle z|\psi\rangle|^2 .
\end{equation}

The main comparison uses the representative dense $n=10$ instance at $p=3$ with a random-strength schedule. At every point in the strength--imbalance scan, $P_{\rm tail}(0.9)$ and the native-gradient response $G_{\rm RMS}$, defined in Eq.~\eqref{eq:gradient-rms}, are evaluated from the same final state.

To compare solution quality with local gradient magnitude without assuming a linear relation, all grid points are ranked by $G_{\rm RMS}$ and divided
into ten equal percentile bins. Within each bin, the median $P_{\rm tail}(0.9)$ is reported together with its interquartile range.

Additional robustness tests repeat the solution-quality scan across uniform and random-strength schedules, $p=3$ and $p=10$, and unweighted
and generic real-weighted cost Hamiltonians.  The weak-drive onset of enhanced near-optimal sampling for uniform schedules is derived below.


\subsection{Perturbative onset of near-optimal sampling}
\label{subsec:perturbative-onset}

We derive the weak-drive expansion used in
Eqs.~\eqref{eq:ptail_weak}--\eqref{eq:Sminus}. Define
\begin{equation}
\bar H_c:=\frac{H_c}{\|H_c\|},
\qquad
\bar H_m:=\frac{H_m}{\|H_m\|},
\end{equation}
with $|\psi_0\rangle=|+\rangle^{\otimes n}$.

For a uniform schedule, $S_\ell=S$ and $\alpha_\ell=\alpha$, so the norm-weighted cost and mixer actions reduce to
\begin{equation}
x=S\frac{\alpha}{1+\alpha},
\qquad
y=S\frac{1}{1+\alpha}.
\label{eq:xy_strength_imbalance}
\end{equation}
The depth-$p$ circuit is therefore
\begin{equation}
U_p=
\left(
e^{-iy\bar H_m}
e^{-ix\bar H_c}
\right)^p .
\end{equation}

To obtain the weak-drive expansion, we Taylor-expand one QAOA layer,
$V=e^{-iy\bar H_m}e^{-ix\bar H_c}$, to second order:
\begin{equation}
V=I+K_1+K_2+\mathcal{O}(S^3),
\end{equation}
with
\[
K_1=-i(x\bar H_c+y\bar H_m),
\]
\[
K_2=
-\frac{x^2}{2}\bar H_c^2
-\frac{y^2}{2}\bar H_m^2
-xy\,\bar H_m\bar H_c .
\]
Since $U_p=V^p$, this gives
\begin{equation}
U_p
=
I+pK_1+pK_2
+\frac{p(p-1)}{2}K_1^2
+\mathcal{O}(S^3).
\end{equation}

Let $\Pi_{\mathrm{tail}}$ project onto states with
$C(z)\geq0.9\,C_{\max}$. Then
\[
P_{\mathrm{tail}}(0.9)
=
\langle\psi_0|
U_p^\dagger\Pi_{\mathrm{tail}}U_p
|\psi_0\rangle,
\qquad
P_0=
\langle\psi_0|\Pi_{\mathrm{tail}}|\psi_0\rangle .
\]

Since \([\Pi_{\mathrm{tail}},\bar H_c]=0\) and \(\bar H_m|\psi_0\rangle=|\psi_0\rangle\), the terms linear in \(x\) and \(y\) in the expansion of \(P_{\mathrm{tail}}(0.9)\) vanish. Expanding to second order therefore gives
\begin{equation}
P_{\mathrm{tail}}(0.9)-P_0
=
p(p+1)\,xy\,D_G
+\mathcal{O}(S^3),
\label{eq:ptail_xy_general}
\end{equation}
where \(D_G\) is a graph-dependent coefficient defined by
\begin{equation}
D_G
:=
\langle\psi_0|
\bar H_c\Pi_{\mathrm{tail}}
|\psi_0\rangle
-
\operatorname{Re}
\langle\psi_0|
\bar H_c\bar H_m\Pi_{\mathrm{tail}}
|\psi_0\rangle .
\label{eq:DG_definition}
\end{equation}
Rewriting \(xy\) in terms of \(S\) and \(\alpha\) gives Eq.~\eqref{eq:ptail_weak}, with \[A_{G,p}=p(p+1)D_G.\]

Repeating the same order-by-order expansion for the fixed-ratio nonuniform schedules gives
\begin{equation}
\begin{aligned}
&P_{\mathrm{tail}}(0.9)-P_0 =
\\
&
D_G\frac{\alpha}{(1+\alpha)^2}
\left[
\left(\sum_{\ell=1}^{p}S_\ell\right)^2
+
\sum_{\ell=1}^{p}S_\ell^2
\right]
+\mathcal{O}(\bar S^3).
\end{aligned}
\label{eq:ptail_nonuniform_second_order}
\end{equation}

For profiles normalized by $\sum_\ell S_\ell=p\bar S$, the uniform limit $S_\ell=\bar S$ reduces the bracket to
$p(p+1)\bar S^2$, recovering Eq.~\eqref{eq:ptail_xy_general}.

The leading-order result in Eq.~\eqref{eq:ptail_xy_general} is
consistent with the general bitstring-resolved small-angle expansion of Hadfield \textit{et al.}~\cite{Hadfield2023}; summing their
QAOA$_p$ probability formula over the sector
$C(z)\geq 0.9C_{\max}$ yields the same quadratic contribution.

For the unweighted MaxCut Hamiltonian considered here, the initial state is uniform over all computational-basis states, giving
\[
P_0=\frac{N_{\mathrm{tail}}}{2^n},
\qquad
\mu_{\mathrm{tail}}
=
\frac{1}{N_{\mathrm{tail}}}
\sum_{z:\,C(z)\geq0.9C_{\max}} C(z),
\]
where $N_{\mathrm{tail}}$ is the number of states in the high-quality sector. Hence
\[
\langle\psi_0|
\bar H_c\Pi_{\mathrm{tail}}
|\psi_0\rangle
=
\frac{P_0\mu_{\mathrm{tail}}}{C_{\max}} .
\]

For a bit string $z$, let $z^{(i)}$ denote the string obtained by flipping vertex $i$. For an unweighted graph,
\begin{equation}
\sum_{i=1}^{n} C\!\left(z^{(i)}\right)
=
(n-4)C(z)+2|E|.
\label{eq:single_flip_cut_sum}
\end{equation}
Using $\bar H_m=n^{-1}\sum_i X_i$ then gives
\begin{equation}
\begin{aligned}
&\operatorname{Re}
\langle\psi_0|
\bar H_c\bar H_m\Pi_{\mathrm{tail}}
|\psi_0\rangle
=\\
&
\frac{P_0}{nC_{\max}}
\left[
(n-4)\mu_{\mathrm{tail}}+2|E|
\right].
\end{aligned}
\end{equation}
Therefore
\begin{equation}
\begin{aligned}
D_G
&=
\frac{4P_0}{nC_{\max}}
\left(
\mu_{\mathrm{tail}}-\frac{|E|}{2}
\right),
\\
A_{G,p}
&=
\frac{4p(p+1)}{nC_{\max}}
P_0
\left(
\mu_{\mathrm{tail}}-\frac{|E|}{2}
\right).
\end{aligned}
\label{eq:DG_maxcut}
\end{equation}
This recovers Eq.~\eqref{eq:Agp}. For the positive real-weighted MaxCut instances used in the perturbative validation, the same result holds with $|E|$ replaced by the total edge weight $W=\sum_{(i,j)\in E}w_{ij}$.

Finally, setting $P_{\mathrm{tail}}(0.9)=P_\star>P_0$ and retaining the leading nonvanishing term gives Eq.~\eqref{eq:Sminus}. This expression describes the weak-drive onset of enhanced near-optimal sampling; at larger strength, higher-order and schedule-dependent interference effects become relevant.

\paragraph{Numerical validation.}
We tested the weak-drive prediction using exact state-vector simulations of uniform and random-strength schedules on independent dense and sparse unweighted MaxCut instances with $n=10$, $12$, and $13$, depths
$p=2$, $3$, $4$, and $5$, together with positive real-weighted controls. Targets were parameterized as
$P_\star=P_0+\eta(1-P_0)$, and the first strength at which $P_{\mathrm{tail}}(0.9)$ reached $P_\star$ was determined numerically.

Uniform schedules were compared with Eq.~\eqref{eq:Sminus}, random-strength
schedules with Eq.~\eqref{eq:ptail_nonuniform_second_order}, and weighted controls with the corresponding weighted coefficient. We additionally verified the small-strength quadratic coefficient and the $[p(p+1)]^{-1/2}$ depth dependence for uniform schedules. Agreement was quantified by
$|S_-^{\mathrm{num}}-S_-^{\mathrm{pred}}|/S_-^{\mathrm{pred}}$. As expected, agreement is strongest within the weak-drive regime and decreases as the target moves further into the finite-strength regime. Detailed instance counts, target probabilities, and error statistics are given in Supplementary Sec.~I\,D.

\paragraph{Sampling-based approximation.}

Direct evaluation of Eq.~\eqref{eq:Sminus} requires problem-specific quantities, including \(C_{\max}\) and the conditional tail mean \(\mu_{\mathrm{tail}}\), which are generally not known a priori.
We therefore use a sampling-based surrogate in which the high-quality sector is defined by the upper fraction \(\rho\) of the random-cut distribution. From \(M\) uniformly sampled bit strings, we retain the top-\(\rho\) fraction according to \(C(z)\) and denote their mean cut value by \(\widehat{\mu}_{\rho}\).

Using a relative enhancement target for the surrogate sector,
\begin{equation}
P_\star^{(\rho)}=(1+\delta)\rho ,
\end{equation}
the baseline probability cancels at leading order. For a uniform schedule, the resulting native-parameter boundary is
\begin{equation}
(\gamma\beta)_-
\simeq
\frac{\delta}
{4p(p+1)
\left(\widehat{\mu}_{\rho}-W/2\right)},
\label{eq:quantile_native_boundary}
\end{equation}
where $W=\sum_{(i,j)\in E}w_{ij}$, with $W=|E|$ for unweighted MaxCut.
Equation~\eqref{eq:quantile_native_boundary} therefore requires only random-cut samples and the graph weights. Mapping this boundary to the
norm-weighted strength $S$ additionally requires
$\|H_c\|=C_{\max}$.

\subsection{Entanglement diagnostics}
\label{subsec:methods-entanglement}

State spreading is quantified by the bipartite von Neumann entropy,
\begin{equation}
S_A=-\mathrm{Tr}(\rho_A\log_2\rho_A),
\end{equation}
normalized by the Page entropy for the corresponding balanced bipartition.

For the regime-boundary analysis, uniform $p=3$ circuits are evaluated over the strength--imbalance plane for dense real-weighted MaxCut instances. At each fixed $\alpha$, the threshold $S^\star(\alpha;e)$ is defined as the first strength at which $S_A/S_{\rm Page}$ reaches $e\in\{0.2,0.5,0.8\}$. The crossing is obtained after three-point smoothing of the sampled entropy curve followed by linear interpolation between adjacent strength values.

The family-wide analysis in Supplementary Fig.~13 includes $n=10$, $13$, and $16$, using independent graph, edge-weight, and balanced-bipartition
realizations where available. Threshold curves at fixed  $(n,\alpha,e)$  are summarized by their median and interquartile range. 
The single-instance landscape in Supplementary Fig.~12 illustrates the threshold construction for one \(n=10\) weighted instance and one balanced bipartition.

\subsection{Native preimages of solution regions}
\label{subsec:methods-solution-preimage}

To compare the native accessibility of near-optimal solution regions with system size, uniform $p=3$ landscapes are evaluated for representative dense, unweighted MaxCut instances at $n=10$, $13$, and $16$. The regions

\begin{equation}
\mathcal{R}_{P_\star}
=
\left\{
(S,\alpha):
P_{\rm tail}(0.9)\geq P_\star
\right\},
\quad 
P_\star\in\{0.3,0.5\},
\end{equation}

are defined using fixed probability thresholds so that their extent is free to grow or shrink with system size.

The selected grid cells are mapped to the native $(\gamma,\beta)$ plane using the strength--imbalance transformation. Their native area is obtained by integrating the Jacobian in Eq.~\eqref{eq:size-jacobian} over the selected cells, thereby avoiding assignment of area to empty regions between disconnected components. In Fig.~\ref{fig:solution-preimage}, the direct native preimages are shown for $P_\star=0.5$, while the Jacobian-integrated area is reported for both thresholds.

\subsection{Selection and mapping of responsive regions}
\label{subsec:methods-preimage-selection}

For the gradient-based preimage analysis, the first prominent high-response region is selected within the fixed window
\[
8\leq S_\ell\leq30,
\qquad
0.05\leq\alpha_\ell\leq5.
\]
Within this window, the gradient landscape is reduced to the one-dimensional profile
\[
g(S_\ell)
=
\max_{\alpha_\ell}
\|\nabla_{\gamma,\beta}C\|,
\]
and the first maximum of the smoothed profile is used to locate the early-response band.

Grid points above the 85th percentile of the gradient values within the search window are retained, and connected components are identified using the same rule for every graph and system size. The component associated with the first response peak is then mapped pointwise to the native $(\gamma,\beta)$ plane.

Its extent is summarized by
\[
\Delta\gamma=\max\gamma-\min\gamma,
\qquad
\Delta\beta=\max\beta-\min\beta,
\]
with $\Delta\gamma\,\Delta\beta$ used as a bounding-box area proxy. The same selection window, percentile threshold, smoothing procedure, and connectivity rule are applied throughout without manual adjustment. The resulting native-width and area diagnostics are shown in Supplementary Fig.~7.

\subsection{Weighted--unweighted comparison and strong-drive analysis}
\label{subsec:methods-strong-drive}

Generic real-weighted controls are constructed by assigning independent edge weights
$w_{ij}\sim\mathrm{Uniform}(0.5,1.5)$
to the same underlying graph topology. The mixer Hamiltonian is unchanged, whereas the cost Hamiltonian and its norm are recomputed for the weighted instance. These controls are used to distinguish structures associated with the common spectral periodicity arising from the integer-valued MaxCut spectrum from more general strong-drive behavior.

Strong-drive scans are evaluated directly in the strength--imbalance plane and mapped to the corresponding native parameters without reducing them to a fundamental periodic interval. For unweighted MaxCut, periodically equivalent native points are generated using the exact cost-parameter period
\[
\gamma_\ell\rightarrow\gamma_\ell+2\pi k,
\qquad k\in\mathbb{Z},
\]
and then mapped back into $(S_\ell,\alpha_\ell)$ coordinates. Constant $\gamma_\ell$ curves are obtained from the same transformation and used as a reference for the curved recurrence structure.  The weighted--unweighted strong-drive comparison is shown in Fig.~\ref{fig:strong_drive}. The corresponding periodic images and constant-$\gamma_\ell$ structure are shown in Supplementary Fig.~8, while the resolved weighted--unweighted recurrence comparison is shown in Supplementary Fig.~9. State spreading at strong drive is characterized by the normalized half-system entanglement entropy and the computational-basis participation fraction. 
The latter is defined as
\[
f_{\mathrm{part}}
=
\frac{1}
{2^n\sum_z |\langle z|\psi\rangle|^4}.
\]
Weighted--unweighted comparisons use the same graph topology and mixer Hamiltonian, while depth- and size-robustness tests repeat the real-weighted calculation across the stated values of $p$ and $n$.











\section*{Acknowledgements}
We thank Gabriele Compostella for insightful discussions.




\bibliographystyle{naturemag}
\bibliography{ref}

% ...



\end{document}


\title{Dynamical regimes of QAOA gradient response}  

\author{Zarin Shakibaei}
\affiliation{Technische Universit\"at Berlin, Institute of Mathematics, 10623 Berlin, Germany}
\author{Alexander Schnell}
\affiliation{Technische Universit\"at Berlin, Institute for Physics and Astronomy, 10623 Berlin, Germany}


\vspace{1em}

\section{Supplementary numerical results}
\label{sec:supp-numerics}

\subsection{Depth and schedule robustness of the gradient landscape}
\label{subsec:supp-schedule-depth}

\begin{figure}[!htbp]
    \centering
    \includegraphics[width=\linewidth]
    {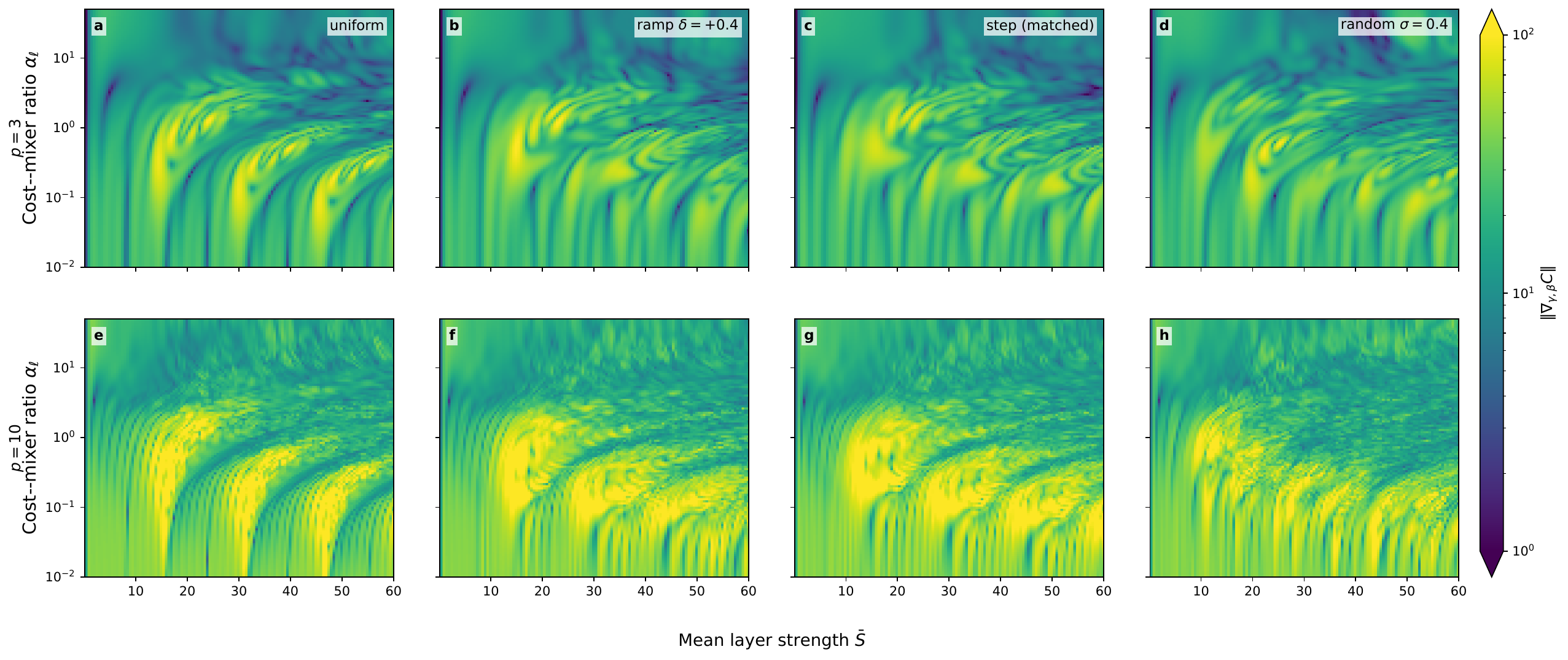}
    \caption{\textbf{Gradient landscapes across depth and schedule shape.}
    Native-gradient response for $p=3$ (top row) and $p=10$ (bottom row). Columns show uniform, ramp, matched-step, and random-strength schedules.
    Nonuniform schedules vary $S_\ell$ at fixed $\alpha_\ell=\alpha$ and satisfy $\sum_\ell S_\ell=p\bar S$.}
    \label{figS:schedule_depth_grid}
\end{figure}

Supplementary Fig.~\ref{figS:schedule_depth_grid} shows the complete comparison of uniform, ramp, matched-step, and random-strength schedules at $p=3$ and $p=10$. For the nonuniform schedules, $S_\ell$ is redistributed
across layers at fixed $\alpha_\ell=\alpha$ and normalized so that $\sum_\ell S_\ell=p\bar S$. Depth and schedule shape substantially alter the fine interference pattern, while the broader locations of suppressed
and enhanced response remain recognizable, providing the landscape-level counterpart to the correlation summary in the main text.

\subsection{Gradient organization under fully random layerwise schedules}
\label{subsec:supp-fully-random-gradient}

To test whether the coarse dynamical organization persists beyond fixed schedule families, we allow both $S_\ell$ and $\alpha_\ell$ to vary across layers while
fixing the global coordinates $(\bar S,\alpha_{\rm tot})$, using the construction defined in Sec.~4.6. This yields a two-dimensional ensemble in which each realization has
the same total norm-weighted cost and mixer actions, while those actions are distributed differently across layers.

\begin{figure}[t]
    \centering
    \includegraphics[width=0.9\textwidth]
    {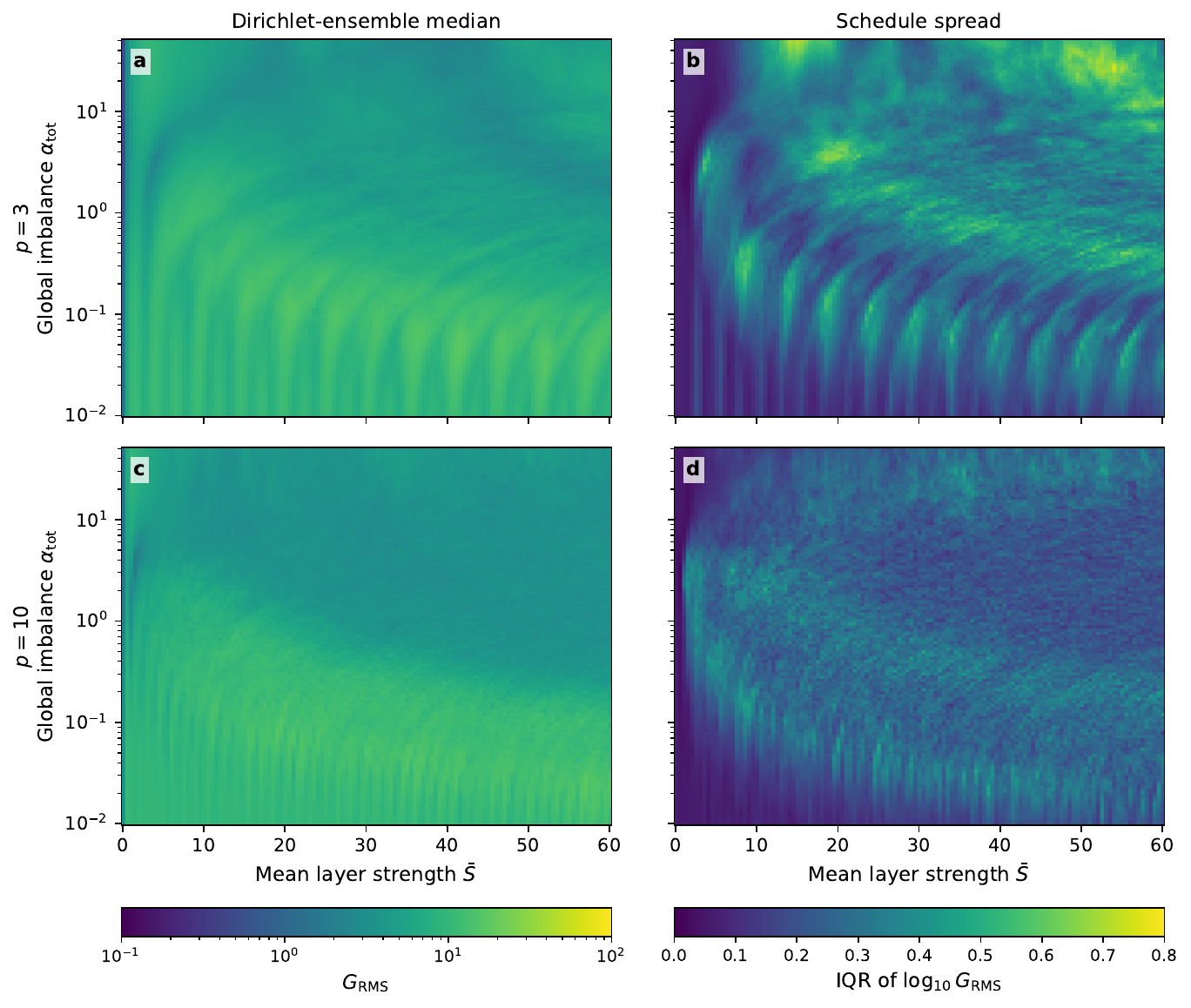}
    \caption{\textbf{Gradient response under fully random layerwise schedules.}
    Ensemble statistics for a representative dense, unweighted $n=10$
    MaxCut instance at $p=3$ (top row) and $p=10$ (bottom row).
    Left panels show the median $G_{\rm RMS}$ at fixed $(\bar S,\alpha_{\rm tot})$; right panels show the corresponding
    interquartile range of $\log_{10}G_{\rm RMS}$. Independent Dirichlet allocations with $\kappa=1$ vary both $S_\ell$ and $\alpha_\ell$ while keeping the global coordinates fixed exactly.}
    \label{fig:supp-fully-random-gradient}
\end{figure}

Supplementary Fig.~\ref{fig:supp-fully-random-gradient} shows that randomizing the layerwise schedule suppresses much of the fine interference structure, while the broad dependence of the gradient response on $(\bar S,\alpha_{\rm tot})$ remains visible at both $p=3$ and $p=10$. The right-hand panels show the interquartile range of $\log_{10}G_{\rm RMS}$ across schedules. The substantial spread, particularly at $p=3$, shows that identical global coordinates can still produce noticeably
different gradient responses. Thus, $(\bar S,\alpha_{\rm tot})$ captures the broad organization of the typical response without uniquely determining an individual schedule's response.

\subsection{Solution quality across schedule, depth, and cost-spectrum controls}
\label{subsec:supp-solution-quality-controls}

\begin{figure}[!htbp]
    \centering
    \includegraphics[width=0.90\linewidth]
    {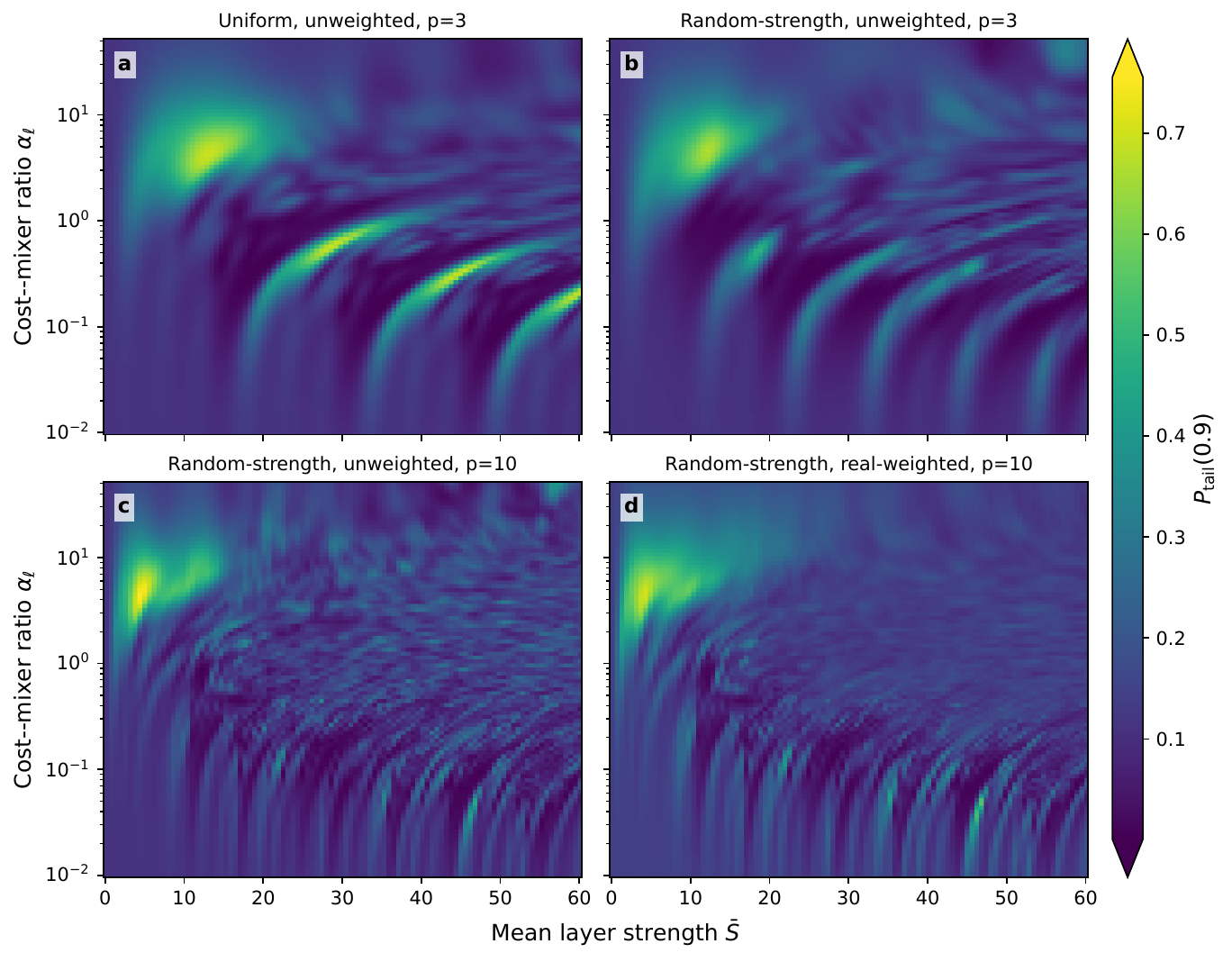}
    \caption{\textbf{Solution-quality landscapes across schedule, depth,
    and cost-spectrum controls.}
    Near-optimal sampling probability $P_{\rm tail}(0.9)$ for a representative dense $n=10$ MaxCut instance, parameterized by mean layer strength $\bar S$ and the common layer ratio $\alpha_\ell=\alpha$.
    \textbf{a}, Unweighted uniform schedule, $p=3$.
    \textbf{b}, Unweighted random-strength schedule, $p=3$.
    \textbf{c}, Unweighted random-strength schedule, $p=10$.
    \textbf{d}, Generic real-weighted random-strength schedule, $p=10$. A common color scale is used across all panels.}
    \label{figS:quality-controls}
\end{figure}

Supplementary Fig.~\ref{figS:quality-controls} tests whether the near-optimal solution region identified in the main text persists under changes in schedule structure, circuit depth, and cost spectrum. We compare uniform and random-strength schedules at $p=3$ and $p=10$, together with unweighted and generic real-weighted MaxCut instances. For the random-strength schedules, $S_\ell$ is redistributed across layers at fixed $\alpha_\ell=\alpha$ and normalized so that $\sum_\ell S_\ell=p\bar S$. Across the cases considered, large $P_{\rm tail}(0.9)$ remains localized within a broadly similar
intermediate-to-cost-biased region of the strength--imbalance plane, although its detailed structure and preferred location vary between cases.

\subsection{Numerical validation of the perturbative onset prediction}
\label{subsec:perturbative-validation}

We numerically validate the perturbative onset prediction using exact state-vector simulations across dense and connected sparse unweighted MaxCut instances with $2n$ edges, together with dense instances having positive real edge weights, for $n=10$, $12$, and $13$ and depths $p=2,3,4,5$. Uniform schedules test the analytical prediction for $S_-$ directly, while random-strength schedules are compared with the corresponding second-order nonuniform expression. We also directly verify the leading quadratic coefficient in the small-\(S\) expansion; across all coefficient tests, the median relative error is of order \(10^{-5}\).

For the onset test, we define the target probability as
$P_\star=P_0+\eta(1-P_0)$ and use $\eta\in\{0.01,0.02,0.05,0.10,0.15\}$. The first numerical crossing is compared with the perturbative prediction, with median relative errors summarized in Supplementary Table~\ref{S-tab:onset-validation}. Agreement is strongest for the smallest target enrichments and decreases as the crossing moves to larger strength, consistent with increasing contributions from the neglected $\mathcal{O}(S^3)$ and higher-order terms. The predicted depth dependence, $S_-\propto[p(p+1)]^{-1/2}$, is also reproduced closely across the tested depths.

\begin{table*}[t]
\centering
\caption{
Validation of the perturbative onset prediction.
For each target increase $\eta$, defined by
$P_\star=P_0+\eta(1-P_0)$, the table reports the median absolute relative error in the predicted onset strength, with the interquartile range in brackets, together with the fraction of cases for which the
target probability is reached within the numerical scan. Errors are evaluated only for cases with a numerical crossing.
}
\label{S-tab:onset-validation}

\renewcommand{\arraystretch}{1.10}
\setlength{\tabcolsep}{7pt}

\begin{tabular}{c l cc cc}
\toprule
& &
\multicolumn{2}{c}{Uniform} &
\multicolumn{2}{c}{Random strength} \\
\cmidrule(lr){3-4}
\cmidrule(lr){5-6}

$\eta$ &
Instance class &
Error [\%] &
Crossing found [\%] &
Error [\%] &
Crossing found [\%] \\
\midrule

\multirow{3}{*}{0.01}
& Dense unweighted
& $0.50\,[0.30,1.08]$ & 100.0
& $0.50\,[0.30,1.09]$ & 100.0 \\
& Sparse unweighted
& $4.82\,[2.22,11.48]$ & 95.0
& $4.92\,[2.32,11.59]$ & 99.6 \\
& Dense real-weighted 
& $0.91\,[0.51,1.51]$ & 100.0
& $0.91\,[0.55,1.51]$ & 100.0 \\

\addlinespace[3pt]

\multirow{3}{*}{0.02}
& Dense unweighted
& $1.00\,[0.55,2.13]$ & 100.0
& $0.97\,[0.55,2.15]$ & 100.0 \\
& Sparse unweighted
& $6.69\,[3.90,16.34]$ & 88.0
& $7.70\,[4.11,17.72]$ & 98.6 \\
& Dense real-weighted 
& $1.75\,[1.05,2.85]$ & 100.0
& $1.76\,[1.07,2.90]$ & 100.0 \\

\addlinespace[3pt]

\multirow{3}{*}{0.05}
& Dense unweighted
& $2.15\,[1.29,4.97]$ & 96.7
& $2.37\,[1.24,5.02]$ & 99.6 \\
& Sparse unweighted
& $13.98\,[6.88,25.22]$ & 72.7
& $20.68\,[8.28,33.98]$ & 88.9 \\
& Dense real-weighted
& $3.72\,[2.55,5.39]$ & 89.7
& $3.99\,[2.55,8.17]$ & 98.4 \\

\addlinespace[3pt]

\multirow{3}{*}{0.10}
& Dense unweighted
& $3.70\,[1.51,6.13]$ & 80.7
& $4.30\,[1.91,8.07]$ & 91.7 \\
& Sparse unweighted
& $17.31\,[11.85,28.11]$ & 53.3
& $21.44\,[12.45,41.86]$ & 67.0 \\
& Dense real-weighted 
& $5.64\,[2.93,7.39]$ & 70.3
& $6.34\,[4.05,21.79]$ & 87.7 \\

\addlinespace[3pt]

\multirow{3}{*}{0.15}
& Dense unweighted
& $4.22\,[2.05,7.38]$ & 68.0
& $5.14\,[2.46,13.21]$ & 80.1 \\
& Sparse unweighted
& $16.46\,[11.64,27.94]$ & 43.3
& $18.96\,[11.73,33.20]$ & 46.6 \\
& Dense real-weighted 
& $5.87\,[4.15,7.30]$ & 57.7
& $6.78\,[4.71,10.69]$ & 68.3 \\

\bottomrule
\end{tabular}
\end{table*}

\subsection{Validation of the sampling-based boundary}

We tested the sampling-based approximation across dense and connected sparse unweighted MaxCut instances and positive real-weighted dense instances, using both uniform and random-strength schedules. From uniformly sampled classical bit strings, we identify the top-$\rho$ sector and estimate its conditional mean $\widehat{\mu}_{\rho}$. The resulting second-order prediction is then compared with the first crossing obtained from exact QAOA state-vector evolution.

The approximation converges rapidly with the number of classical samples. For representative upper-tail fractions, $10^3$ random cuts already yield sub-percent median sampling error relative to the corresponding exact quantile-based boundary, with further improvement as the sample size increases. End-to-end comparisons remain accurate in the weak-drive regime across dense, sparse, weighted, and unweighted instances and for both uniform and random-strength schedules. Thus, the perturbative onset for a quantile-defined high-quality sector can be estimated without exact enumeration of the classical cost spectrum.

\subsection{Robustness of the preferred solution-quality cost--mixer ratio}
\label{S-subsec:supp-quality-ratio-robustness}

We assess the robustness of the cost--mixer ratio associated with near-optimal sampling across graph instances, circuit depths, schedule families, cost Hamiltonians, and system sizes. For each landscape, we retain the top $5\%$ of parameter points ranked by $P_{\rm tail}(0.9)$ and define the preferred cost--mixer ratio as the median $\alpha$ within this subset.

\begin{figure}[!htbp]
    \centering
    \includegraphics[width=0.80\linewidth]
    {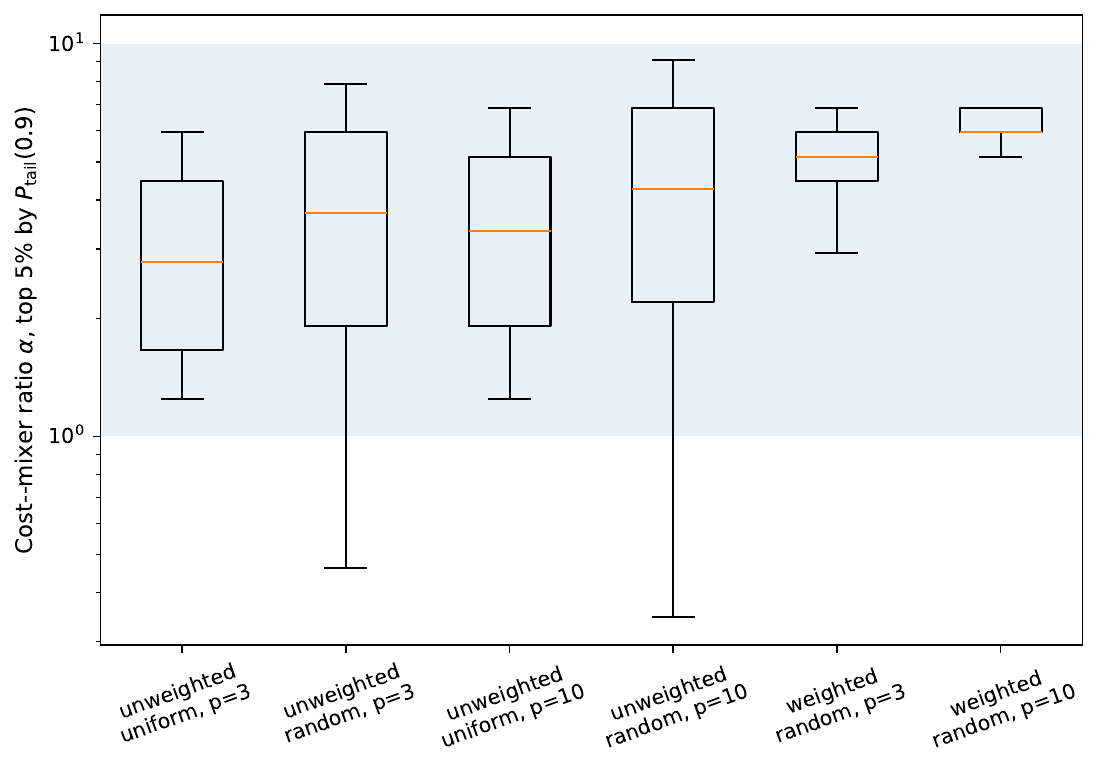}
    \caption{\textbf{Multi-instance robustness of the preferred
    solution-quality cost-mixer ratio.}
    Distribution of the median $\alpha$ among the top $5\%$ of parameter points ranked by $P_{\rm tail}(0.9)$. Each observation corresponds to one independent landscape. Boxes compare unweighted uniform and random-strength schedules at $p=3$ and $p=10$, together with generic real-weighted random-strength schedules. The shaded region marks
    $1\leq\alpha\leq10$.}
    \label{figS:quality-alpha-robustness}
\end{figure}

\begin{figure}[t]
    \centering
    \includegraphics[width=0.8\linewidth]
    {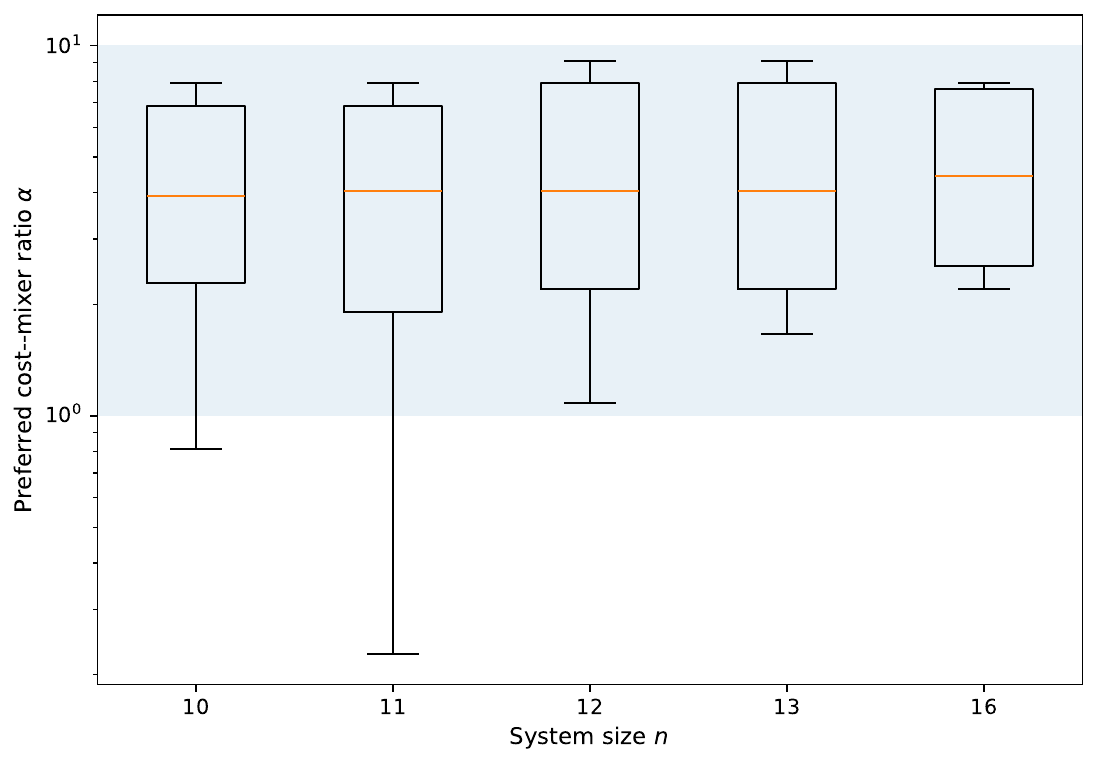}
    \caption{\textbf{System-size robustness of the preferred cost--mixer ratio for near-optimal sampling.}
    Distribution of the preferred cost--mixer ratio $\alpha$ for unweighted MaxCut at $p=10$, defined for each instance--schedule landscape as the median $\alpha$ among the top $5\%$ of grid points ranked by $P_{\rm tail}(0.9)$. Results combine uniform and random-strength schedules across six graph instances for each of $n=10$--$13$ and two instances for $n=16$. The smaller sample at $n=16$ reflects the smaller number of available graph instances.}
    \label{fig:supp-alpha-size-robustness}
\end{figure}

Supplementary Fig.~\ref{figS:quality-alpha-robustness} shows that this preferred ratio remains predominantly within a balanced-to-cost-biased range across uniform and random-strength schedules, depths $p=3$ and $p=10$, and unweighted and generic real-weighted MaxCut instances, while retaining substantial landscape-to-landscape variability. We therefore use $1\leq\alpha\leq10$ as a descriptive range rather than a sharp boundary.
To test whether this preferred ratio drifts systematically with system size, Supplementary Fig.~\ref{fig:supp-alpha-size-robustness} applies the same statistic to unweighted MaxCut landscapes at $p=10$ across $n=10$--$16$. No pronounced systematic drift is observed over the sizes considered, and the preferred ratio remains predominantly within the same balanced-to-cost-biased regime.



\subsection{Solution quality under fully random layerwise schedules}
\label{subsec:supp-fully-random-quality}

\begin{figure}[t]
    \centering
    \includegraphics[width=\textwidth]
    {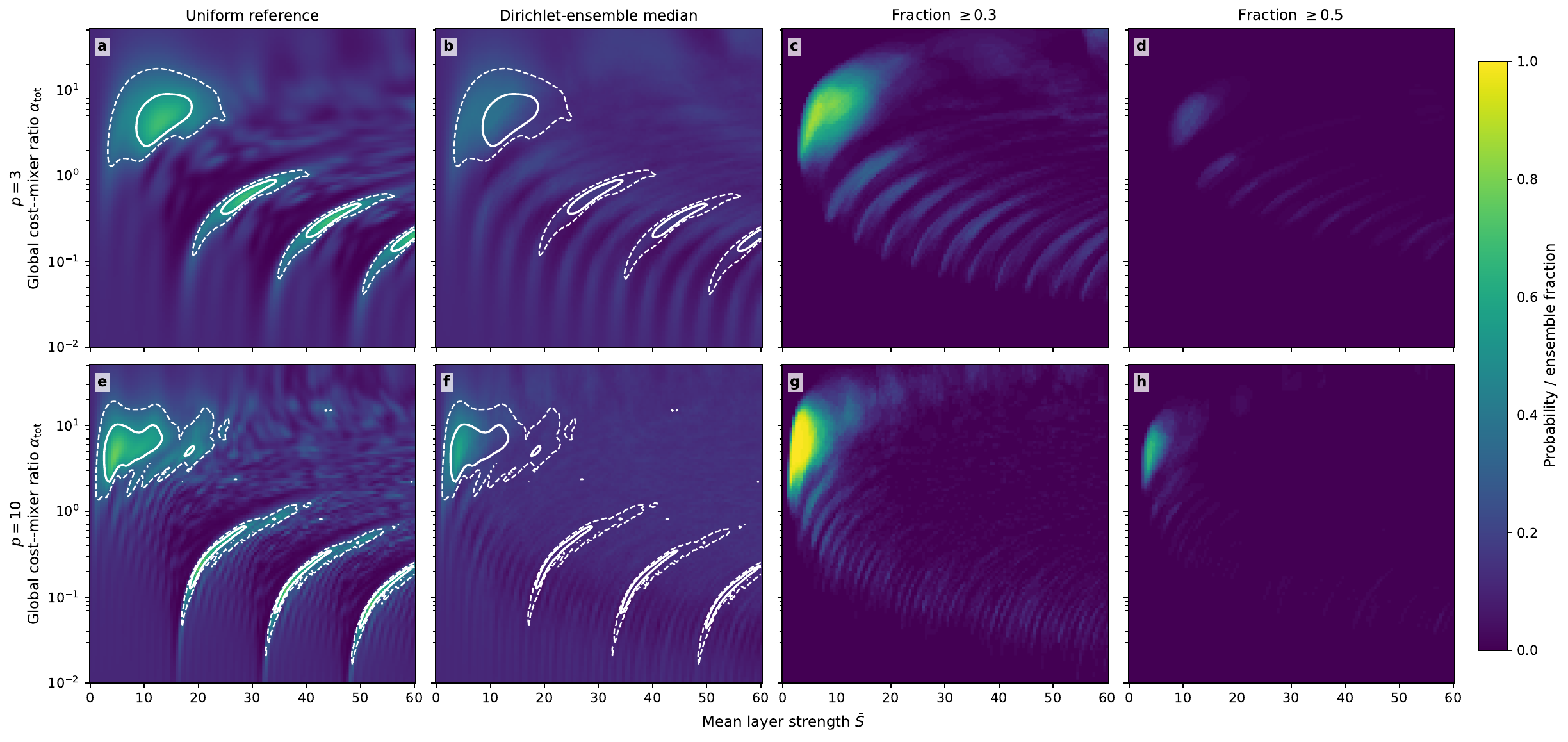}
    \caption{\textbf{Solution quality under fully random layerwise schedules.}
    Results for a representative dense, unweighted $n=10$ MaxCut instance at
    $p=3$ (top row) and $p=10$ (bottom row). From left to right, panels show
    the uniform-schedule $P_{\rm tail}(0.9)$ landscape, the median over the
    fully random Dirichlet ensemble, and the fractions of schedules satisfying
    $P_{\rm tail}(0.9)\geq0.3$ and $P_{\rm tail}(0.9)\geq0.5$.
    White contours in the first two columns mark the $0.3$ and $0.5$ levels
    of the corresponding uniform landscape.}
    \label{fig:supp-fully-random-quality}
\end{figure}

The fully random schedule ensemble introduced in
Supplementary Sec.~\ref{subsec:supp-fully-random-gradient} is used to assess whether the near-optimal solution region persists when both $S_\ell$ and $\alpha_\ell$ vary across layers. Supplementary Fig.~\ref{fig:supp-fully-random-quality} shows that the ensemble median
remains concentrated in approximately the same region of $(\bar S,\alpha_{\rm tot})$ as the uniform reference at both $p=3$ and $p=10$. The fractions of schedules reaching $P_{\rm tail}(0.9)\geq0.3$ and $P_{\rm tail}(0.9)\geq0.5$ further show that near-optimal schedules are statistically concentrated within this region,
although individual schedules at the same global coordinates can perform very differently.

At fixed $(\bar S,\alpha_{\rm tot})$, all sampled schedules have the same total cost and mixer actions and the same total evolution time. Differences in $P_{\rm tail}(0.9)$ at a given point therefore arise from how these actions are distributed across layers.

\subsection{Additional diagnostics of native-preimage contraction}
\label{supp:preimage-diagnostics}

\begin{figure}[!htbp]
    \centering
    \includegraphics[width=\textwidth]
    {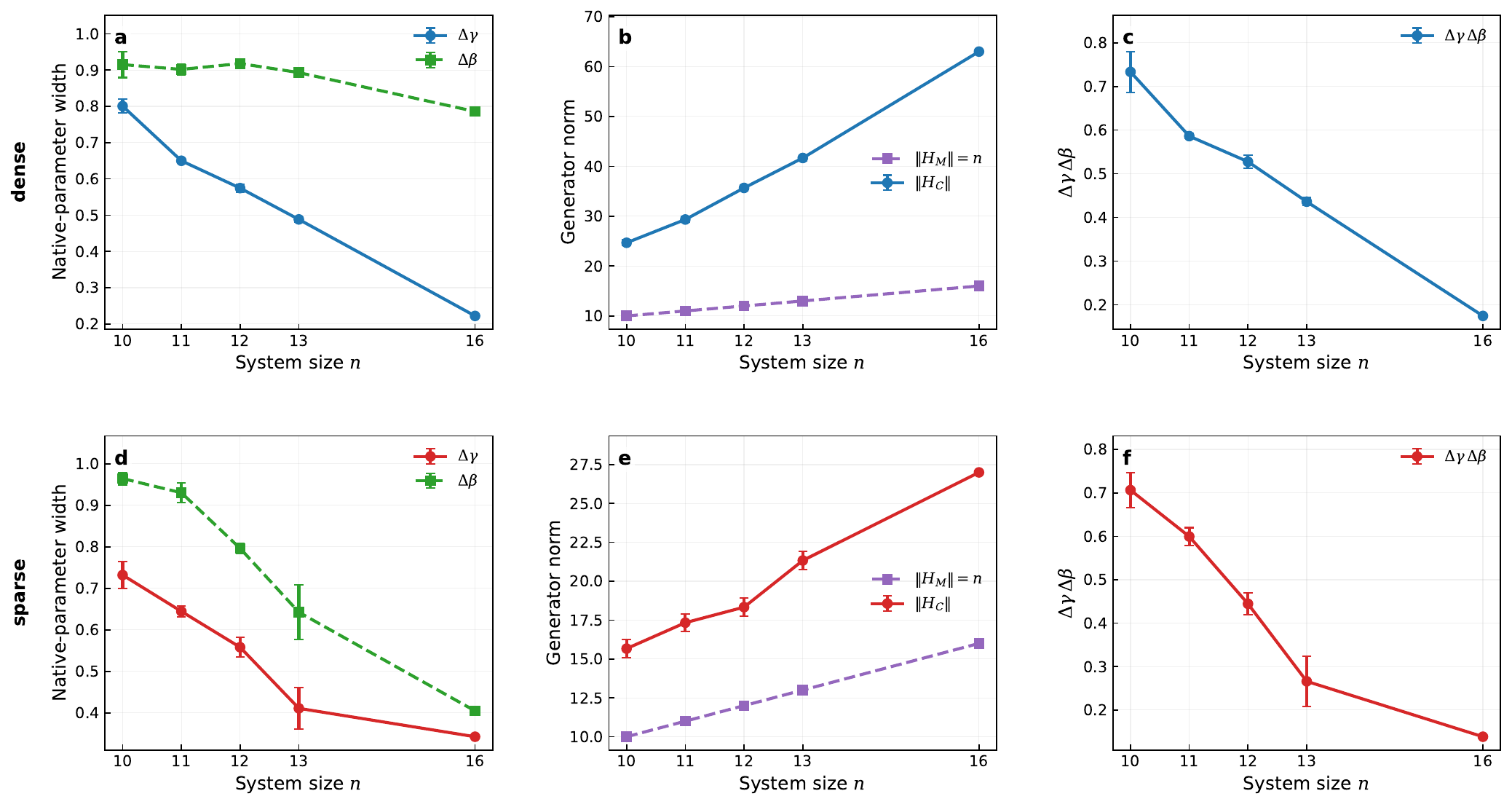}
    \caption{\textbf{Diagnostics of native-preimage contraction for dense
    and sparse graph families.}
    Panels \textbf{a--c} show dense graphs and panels \textbf{d--f} sparse
    connected graphs with $2n$ edges. Panels \textbf{a,d} show the native
    widths $\Delta\gamma$ and $\Delta\beta$, panels \textbf{b,e} the
    generator norms, and panels \textbf{c,f} the area proxy
    $\Delta\gamma\,\Delta\beta$. Points denote means over available graph
    instances and error bars one sample standard deviation; only one instance
    is available at $n=16$.}
    \label{fig:supp-preimage-diagnostics}
\end{figure}

Supplementary Fig.~\ref{fig:supp-preimage-diagnostics} examines how the selected high-response region changes with system size after being mapped from strength--imbalance coordinates into the native $(\gamma,\beta)$ plane. For dense graphs, the native preimage narrows mainly along the cost-parameter direction $\Delta\gamma$, whereas $\Delta\beta$ changes only weakly. For sparse graphs, both $\Delta\gamma$ and $\Delta\beta$ decrease with system size. The corresponding area proxy $\Delta\gamma\,\Delta\beta$ decreases
overall for both graph families as the system size grows.

\subsection{Native periodicity and cost-spectrum recurrence}
\label{subsec:supp-native-periodicity}

\begin{figure}[!htbp]
    \centering
    \includegraphics[width=\textwidth]
    {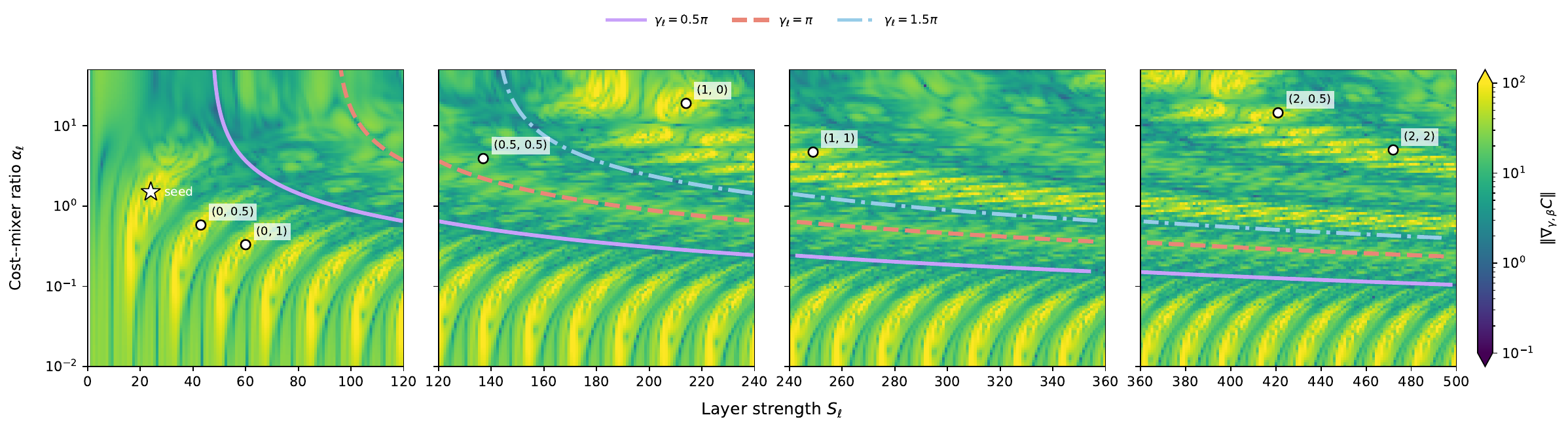}
    \caption{
        \textbf{Native periodicity and constant-$\gamma$ structure
        in the strength--imbalance plane.}
        Native-gradient response over successive strength intervals for an
        unweighted MaxCut instance. The star marks the seed point. A marker
        labelled $(k_\gamma,k_\beta)$ denotes the native shift
        $\gamma=\gamma_{\rm seed}+2\pi k_\gamma$ and
        $\beta=\beta_{\rm seed}+\pi k_\beta$, subsequently mapped into
        $(S,\alpha)$ coordinates. Curves show constant native cost parameters
        $\gamma=0.5\pi$, $\pi$, and $1.5\pi$.}
    \label{figS:periodic-images}
\end{figure}

Supplementary Fig.~\ref{figS:periodic-images} shows how the exact native cost-parameter periodicity of unweighted MaxCut appears after transformation to $(S_\ell,\alpha_\ell)$ coordinates. Periodically equivalent native points map to separated and distorted locations as $S_\ell$ increases, rather than to simple translations along the strength axis. The partial alignment of constant-$\gamma_\ell$ contours with curved low-response valleys further indicates that part of the strong-drive structure reflects the underlying native cost-phase dependence, although the full response also depends on $\beta_\ell$ and the alternating circuit dynamics.

\begin{figure}[!htbp]
    \centering
    \includegraphics[width=0.75\linewidth]
    {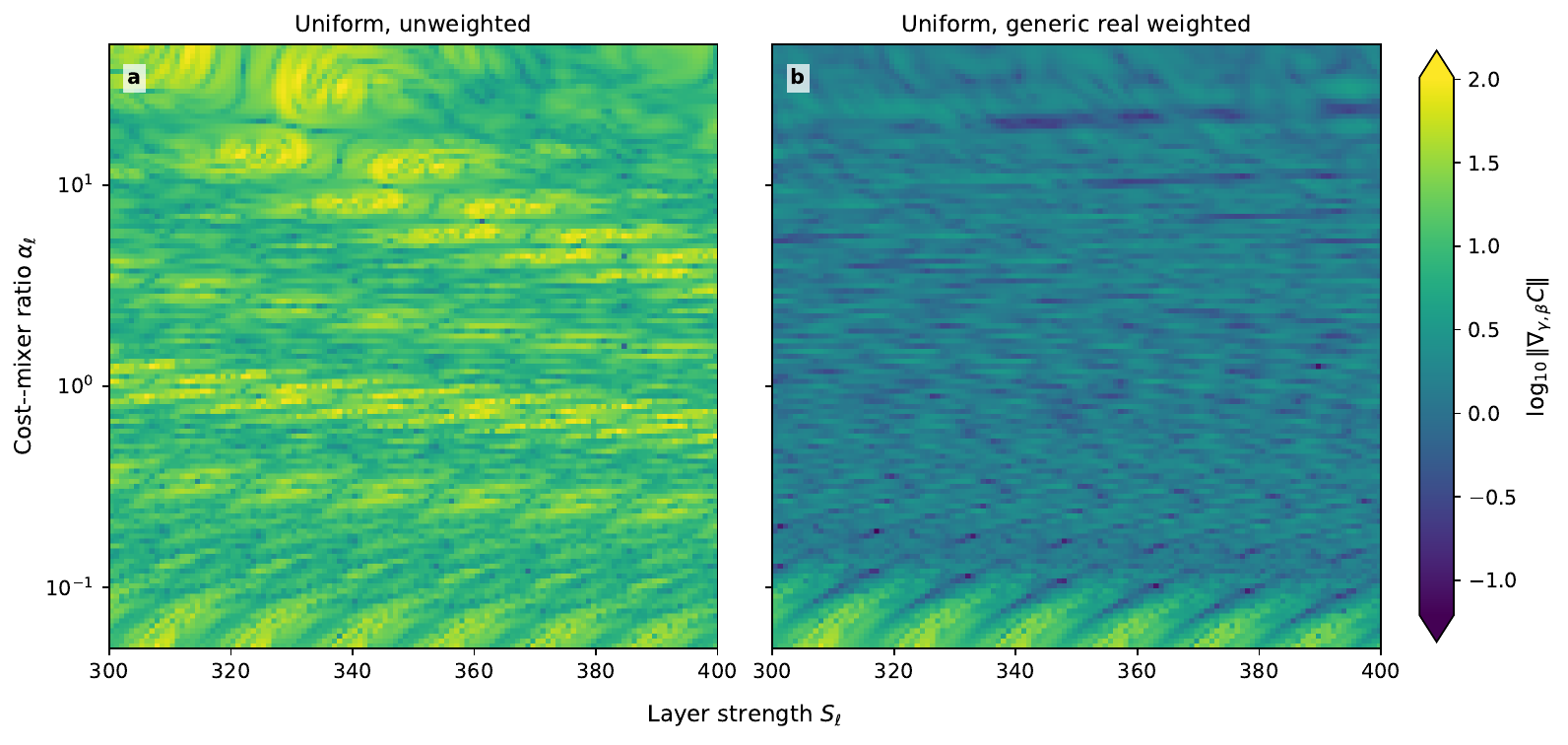}
    \caption{\textbf{Resolved weighted--unweighted recurrence control.}
    Native-gradient response for representative dense $n=10$, $p=3$ MaxCut instances over $S_\ell\in[300,400]$. \textbf{a}, Unweighted MaxCut.
    \textbf{b}, The same graph topology with generic real edge weights. The panels use the same color scale.}
    \label{figS:resolved_recurrence_control}
\end{figure}

To isolate the role of the commensurate cost spectrum, Supplementary Fig.~\ref{figS:resolved_recurrence_control} compares unweighted and generic real-weighted MaxCut over the lower strong-drive window $S_\ell\in[300,400]$. Both landscapes retain high-response structure at low $\alpha_\ell$, consistent with the mixer periodicity common to the two calculations. The additional recurrent high-response features visible in the unweighted landscape are absent in the generic real-weighted case. This contrast supports their association with the commensurate cost spectrum and the resulting exact native cost-phase recurrence of unweighted MaxCut, rather than with strong driving alone.



\subsection{Robustness of strong-drive state spreading}
\label{S-subsec:supp-strong-drive-spreading}

\begin{figure}[!htbp]
    \centering
    \includegraphics[width=0.8\linewidth]
    {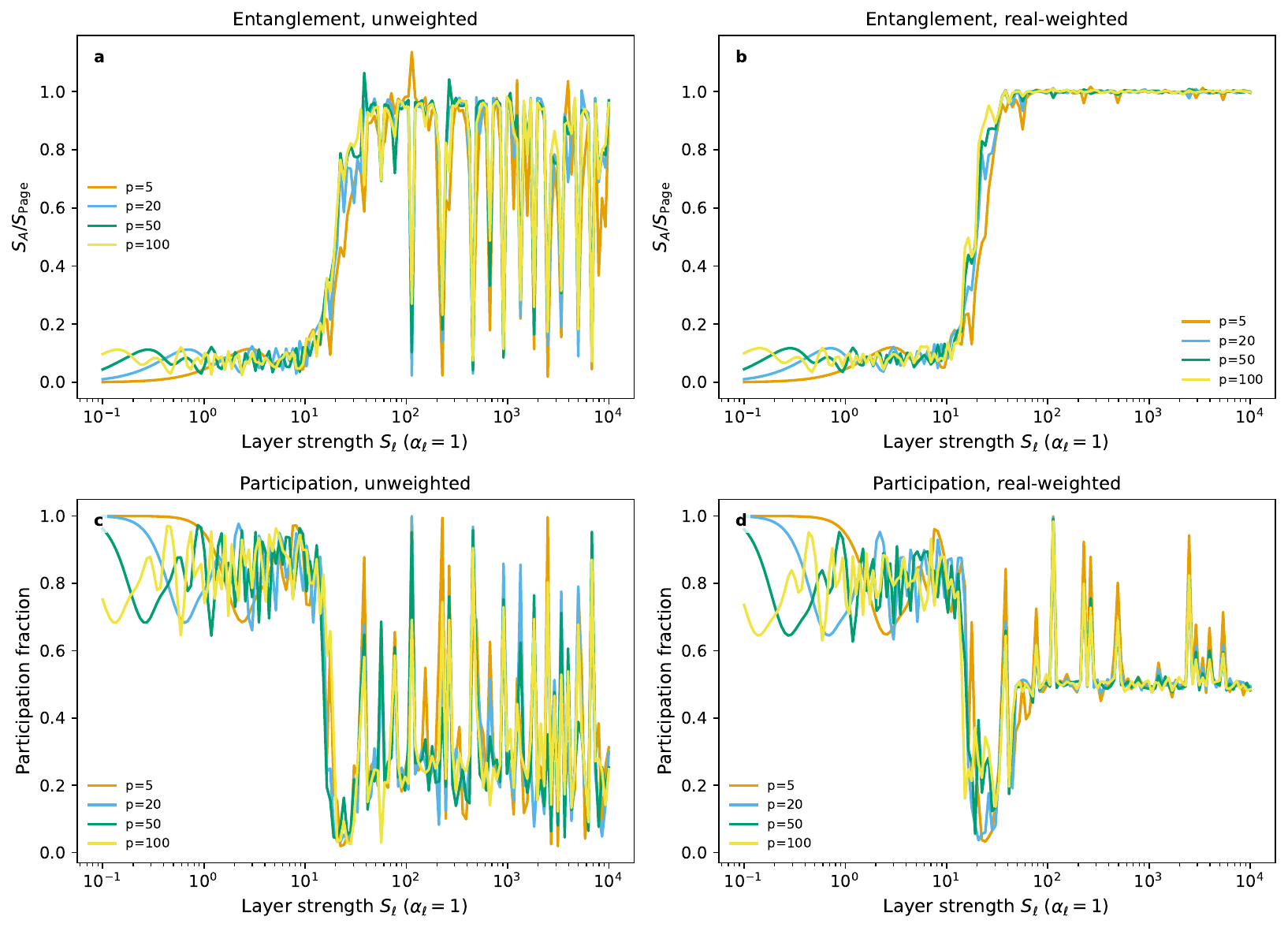}
    \caption{\textbf{Weighted--unweighted strong-drive comparison across
    circuit depth.}
    Normalized half-system entanglement entropy (\textbf{a,b}) and
    computational-basis participation fraction (\textbf{c,d}) for
    representative dense $n=10$ MaxCut instances, comparing unweighted
    (\textbf{a,c}) and generic real-weighted (\textbf{b,d}) cost Hamiltonians
    at $\alpha_\ell=1$ and $p=5,20,50,100$.}
    \label{figS:weighted_unweighted_depth}
\end{figure}

We first test the weighted--unweighted contrast across circuit depth. Supplementary Fig.~\ref{figS:weighted_unweighted_depth} compares normalized half-system entanglement and computational-basis participation for representative dense $n=10$ instances at $p=5$, $20$, $50$, and $100$. At sufficiently large strength, the generic real-weighted evolution remains close to the Page-entangled, participation-$1/2$ regime across all tested depths. The unweighted evolution also reaches highly entangled states but exhibits recurrent entropy collapses and large participation excursions. The distinction identified in the main text is therefore not specific to a single circuit depth.

\begin{figure}[!htbp]
    \centering
    \includegraphics[width=\linewidth]
    {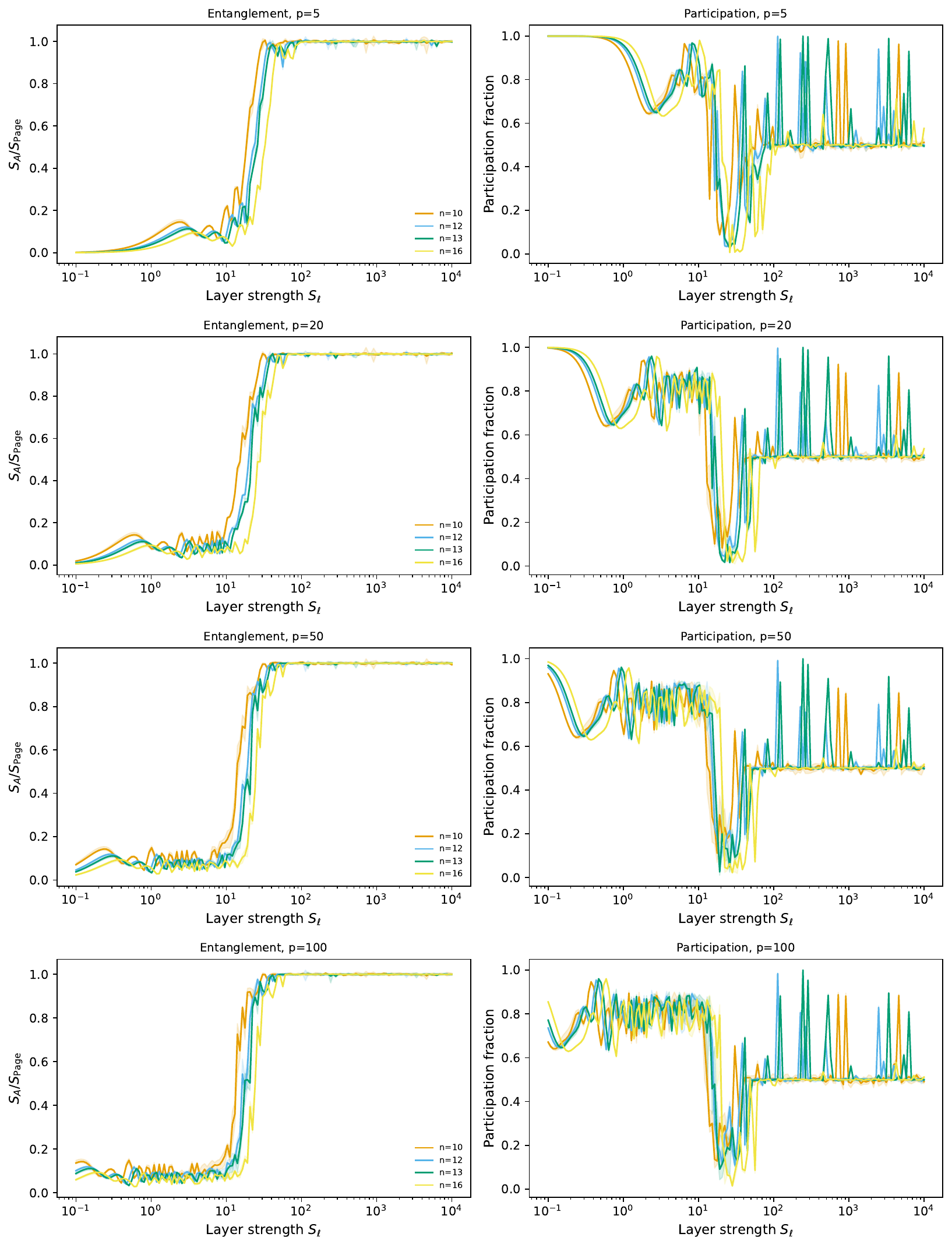}
    \caption{\textbf{Depth- and system-size robustness of strong-drive state
    spreading.}
    Normalized half-system entanglement entropy (left column) and computational-basis participation fraction (right column) for generic
    real-weighted dense MaxCut at $n=10,12,13,16$ and
    $\alpha_\ell=1$. Rows correspond to $p=5,20,50,100$; curves summarize
    independent real-weight realizations.}
    \label{figS:strong_drive_robustness}
\end{figure}

We next test the persistence of the real-weighted strong-drive regime across both system size and depth. Supplementary Fig.~\ref{figS:strong_drive_robustness} shows results for dense real-weighted MaxCut at $n=10$, $12$, $13$, and $16$ and $p=5$, $20$, $50$, and $100$. At sufficiently large $S_\ell$, the normalized half-system entropy approaches the Page value and the participation fraction remains near $1/2$ across the tested $n$ and $p$. The location and intermediate structure of the crossover vary with system size and depth, but the strong-drive state-spreading regime remains qualitatively robust over the range examined.



\subsection{Entanglement crossover across strength and imbalance}
\label{S-subsec:supp-entanglement-boundary}

\begin{figure}[!htbp]
    \centering
    \includegraphics[width=0.72\linewidth]
    {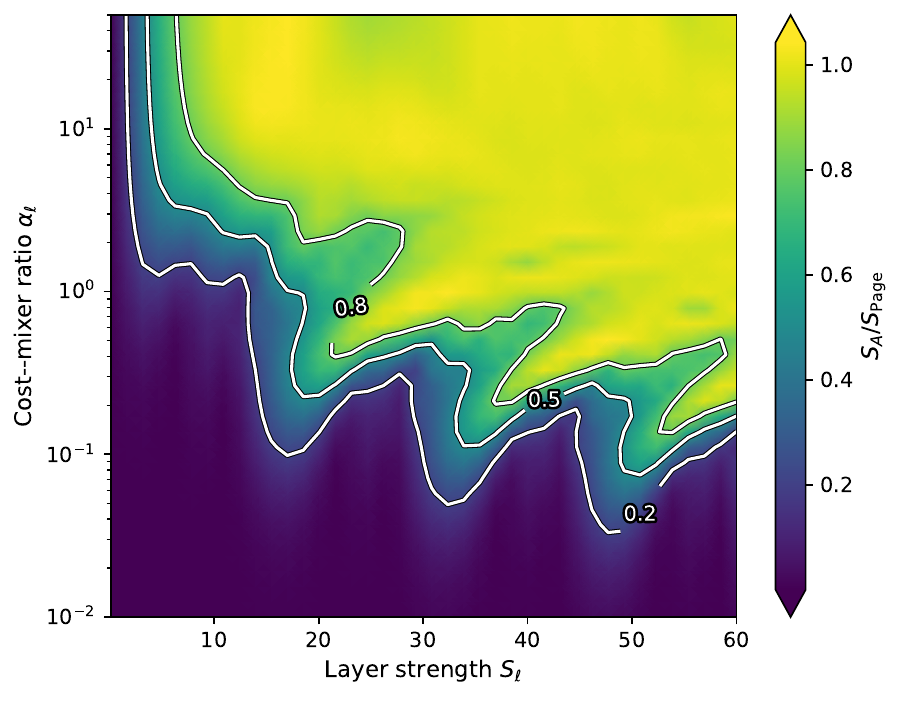}
    \caption{\textbf{Illustrative entanglement landscape and threshold
    construction.}
    Normalized bipartite entanglement entropy $S_A/S_{\rm Page}$ over the
    $(S_\ell,\alpha_\ell)$ plane for a single dense real-weighted $n=10$
    MaxCut instance at $p=3$ and one balanced bipartition. White contours mark
    $S_A/S_{\rm Page}=0.2$, $0.5$, and $0.8$; their first crossings along
    increasing $S_\ell$ define the corresponding threshold strengths.}
    \label{fig:entanglement_landscape}
\end{figure}

Supplementary Fig.~\ref{fig:entanglement_landscape} provides an illustrative example of the entanglement landscape underlying the crossover-boundary construction discussed in the main text. For a single dense real-weighted \(n=10\) MaxCut instance and one balanced bipartition, the normalized bipartite entropy is evaluated over the \((S_\ell,\alpha_\ell)\) plane. At each fixed \(\alpha_\ell\), we define \(S_\ell^\star(\alpha_\ell;e)\) as the smallest layer strength for which \(S_A/S_{\rm Page}\) reaches a target level \(e\). We consider \(e=0.2\), \(0.5\), and \(0.8\).
Applying the same construction across independent graph instances, weight realizations, and balanced bipartitions yields the aggregated crossover curves shown below in 
For dense real-weighted instances at \(p=3\), the threshold strength generally decreases as the dynamics become more cost dominated, while higher target entanglement levels require larger strengths. The detailed crossover varies with system size and becomes less regular in the strongly mixer-dominated regime. We therefore interpret \(S_\ell^\star(\alpha_\ell;e)\) as a finite-size crossover boundary rather than a sharp transition.
Across the tested system sizes, the same qualitative dependence persists: both the overall strength and the cost--mixer imbalance determine where substantial bipartite entanglement first appears.

\begin{figure}
    \centering
    \includegraphics[width=0.60\textwidth]
    {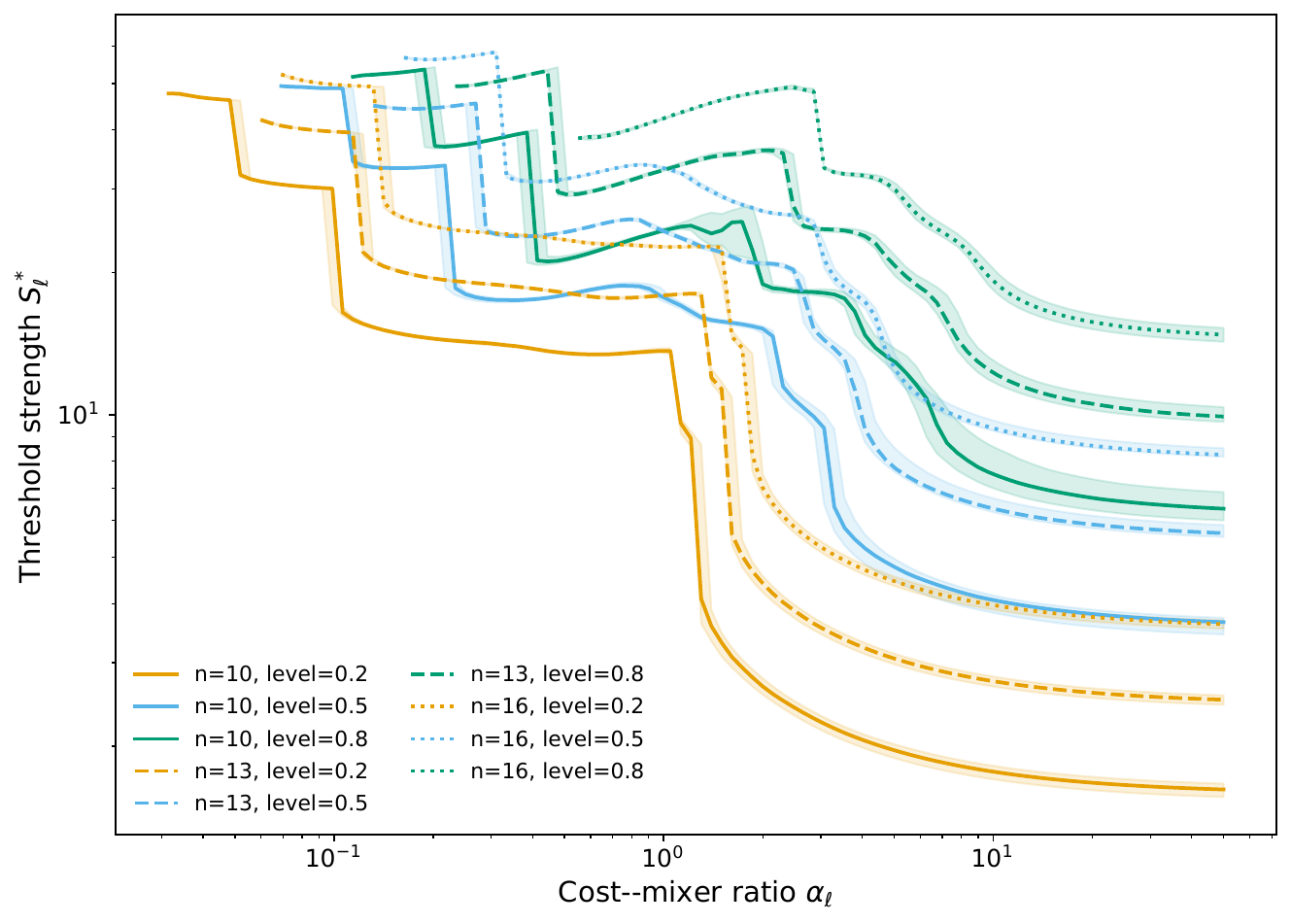}
    \caption{\justifying \textbf{Strength--imbalance organization of the entanglement
    crossover.}
    Threshold strength $S_\ell^\star$ at which the normalized bipartite entropy $S_A/S_{\rm Page}$ first reaches $e=0.2$, $0.5$, and $0.8$ as a function of $\alpha_\ell$ at depth $p=3$ for dense, real-weighted MaxCut instances with $n=10$, $13$, and $16$. Curves show medians over graph
    instances, independent weight realizations, and balanced bipartitions; shaded regions denote the interquartile range. For $n=10$ and $n=13$, three graph variants with five weight realizations each are used; for $n=16$, one graph variant with five weight realizations is used. Three
    balanced bipartitions are evaluated for each weighted realization.}
    \label{fig:entanglement_crossover}
\end{figure}




